\documentclass[twocolumn,prstab,floatfix,draftcopy]{revtex4}
\usepackage{graphicx}
\usepackage{latexsym}
\usepackage{amsfonts}
\usepackage{amssymb}
\usepackage{bm}
\usepackage{amsmath}

\newcommand{\BfE}{{\bf E}}
\newcommand{\BfL}{{\bf L}}
\newcommand{\Bfr}{{\bf r}}
\newcommand{\BfP}{{\bf P}}
\newcommand{\BfJ}{{\bf J}}
\newcommand{\Bfn}{{\bf n}}
\newcommand{\Bfa}{{\bf a}}
\newcommand{\calE}{{\cal E}}
\newcommand{\Csr}{{\mbox{\tiny CSR}}}
\newcommand{\Sc}{{\mbox{\tiny SC}}}
\newcommand{\mb}[1]{{\mbox{\tiny #1}}}
\newcommand{\calO}{{\mathcal O}}
\newcommand{\sx}{\sigma_x}
\newcommand{\sy}{\sigma_y}
\newcommand{\sz}{\sigma_z}

\begin{document}

\title{Extended 1D Method for Coherent Synchrotron Radiation including Shielding}

\author{David C. Sagan, Georg H. Hoffstaetter, Christopher E. Mayes, Udom Sae-Ueng}
\affiliation{Cornell University, Ithaca, New York 14853}

\begin{abstract}
Coherent Synchrotron Radiation can severely limit the performance of
accelerators designed for high brightness and short bunch
length. Examples include light sources based on ERLs or FELs, and
bunch compressors for linear colliders. In order to better simulate
Coherent Synchrotron Radiation, the established 1-dimensional
formalism is extended to work at lower energies, at shorter bunch
lengths, and for an arbitrary configuration of multiple bends. Wide
vacuum chambers are simulated by means of vertical image charges. This
formalism has been implemented in the general beam dynamics code {\tt
Bmad} and its results are here compared to analytical approximations,
to numerical solutions of the Maxwell equations, and to the simulation
code {\tt elegant}.
\end{abstract}

\maketitle

\section{Introduction}

It is envisioned that future accelerators will call for shorter beams
of higher intensity. A possible limiting factor in these efforts is an
increase in energy spread and transverse emittance, as well as a
micro-bunching instability, due to Coherent Synchrotron Radiation
(CSR).

The first CSR calculations were performed by Schwinger in 1945. Using
a Green's function method, he arrives at the power spectrum of a
single charge bending in free space as well as between infinite
conducting plates, and thereby computes the coherent power radiated by
a collection of charges~\cite{schwinger45}. Warnock extends this work
to include the longitudinal impedance on a bunched
beam~\cite{warnock90}. Many papers covering the history and importance
of CSR forces can be found in \cite{murphy04}.

This paper uses an approach to calculate the CSR wake-field
originating with Saldin \emph{et al}.~\cite{saldin97} and generalized by
Sagan\cite{sagan06}. Here we calculate the CSR force between two
charges traveling on the same curve, and integrate over a longitudinal
bunch distribution to give a longitudinal wake-field. Transverse
particle coordinates and transverse force components are
neglected. The formalism developed here is generalized to include
arbitrary lattice configurations of bends and drifts, including, for
example, radiation from one bend entering another, and bend radiation
extending into drift regions.

Simulating CSR effects is the subject of a number of codes. The method
here is implemented in the particle tracking code {\tt
Bmad}~\cite{b:bmad}. Our simulation results are compared with two of
the codes described in \cite{bassi06} and with approximate analytic
formulas.

\section{Two Particle Interaction}

\begin{figure}[bt!]
\includegraphics[width=3.1in,height=1.8in]{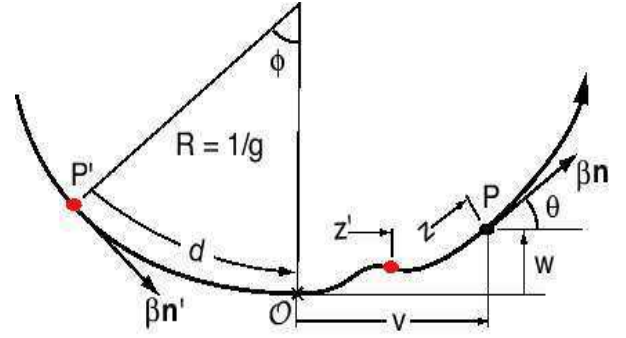}
\caption{A particle at point $\BfP'$ kicks a particle at point
$\BfP$.\label{fg:2part}}
\end{figure}

In order to compare the force acting on one particle from the
radiation that is emitted by another, the analysis starts by
considering two particles of charge $e$ following the same trajectory
as shown in Fig.~\ref{fg:2part}. The Li\'{e}nard-Wiechert formula
\cite{jackson99} gives the electric field $\BfE(\BfP)$ at the position
of the kicked particle at point $\BfP$ and time $t$ due to the source
particle at point $\BfP'$ and retarded time $t'$
\begin{equation}
  {\BfE(\BfP)} = \frac{e}{4 \pi \epsilon_0 }
  \frac{\frac{1}{\gamma^2}(\BfL - L \beta \Bfn') + 
        \frac{1}{c^2}\BfL \times [(\BfL - L \beta \Bfn') \times \Bfa']}
  {(L - \BfL \cdot \beta \Bfn')^3}
  \label{eq:epegnb}
\end{equation}
It will be assumed that both particles have the same constant speed
$\beta = v/c$, $\Bfn'$ and $\Bfn$ are the unit velocity vectors for
the source and kicked particles respectively, and $\BfL = \BfP -
\BfP'$ is the vector from the source point to the kick point. The
retarded time $t'$ is related to $t$ via $t - t' = L/c$.  At time $t$,
the source particle has a longitudinal position $z'$ with respect to
the bunch center and the longitudinal position of the kicked particle
is $z$. The distance $\zeta \equiv z - z'$ between the particles at
constant time can be computed via the equation
\begin{equation}
  \zeta = L_s - \beta \, L\ ,
  \label{eq:sslbl}
\end{equation}
where $L_s$ is the path length from $\BfP'$ to $\BfP$.  Generally, the
relativistic approximation $\beta = 1$ will be made. However, some
terms in $1 - \beta \simeq 1/2 \, \gamma^2$ will need to be retained.

The first term on the right hand side of Eq.~(\ref{eq:epegnb}) has a
$1/\zeta^2$ singularity at small distances.  Following Saldin \emph{et al.}~\cite{saldin97},
this singularity is dealt with by dividing the electric field into two
parts. The space charge component $\BfE_\Sc$, which contains the
singularity, is the field that would result if the particles where
moving without acceleration along a straight line. The CSR term,
$\BfE_\Csr$, is what is left after subtracting off the space charge
term
\begin{equation}
  \BfE_\Sc \equiv \frac{e}{4 \pi \epsilon_0}\frac{\Bfn}{\gamma^2 \, \zeta^2}
  \, , \qquad
  \BfE_\Csr \equiv \BfE - \BfE_\Sc\ .
  \label{eq:epegss}
\end{equation}
The rate $K \equiv d{\cal E}/ds$ at which the kicked particle is changing
energy due to the field of the source particle is
\begin{equation}
  K \equiv K_\Csr + K_\Sc
  = e \, \Bfn \cdot \BfE_\Csr + e \, \Bfn \cdot \BfE_\Sc\ .
 \label{eq:kebe2}
\end{equation}
Following Saldin \emph{et al.}~\cite{saldin97}, the transverse extent of the beam will be
ignored in the calculation of $K_\Csr$. However, the inclusion of the
finite beam size will be needed to remove the singularity in the
calculation of $K_\Sc$ as discussed in Section~\ref{k_sc}.

\section{CSR Calculation}

The source point $\BfP'$ and the kick point $\BfP$ will, in general, not be
within the same lattice element. Because the transverse extent of the
beam is being ignored, all elements will be considered to be either
bends or drifts.

In Fig.~\ref{fg:2part}, $R$ is the bending radius and $g = 1/R$ is the
bending strength of the element that contains the source point
$\BfP'$.  The magnitude of the acceleration is $a' \simeq c^2/R$.
This element ends at point $\calO$.  The angle between $\BfP'$ and
$\calO$ is $\phi$, and $d = R \, \phi$ is the path length between
$\BfP'$ and $\calO$.

Between point $\calO$ and the kick point $\BfP$, $d_i$ is the path
length within the $i^{th}$ element, $i = 1, \ldots, N$, where $N$ is
the number of elements in this region. For the last element, $d_N$ is
the distance from the start of the element to point $\BfP$. For the
$i^{th}$ element, $\phi_i$ is the bend angle, $R_i$ is the bend
radius, and $g_i = 1 / R_i$ is the bend strength. For a drift $\phi_i
= g_i = 0$.

\begin{figure}[tb]
\includegraphics[width=3.1in]{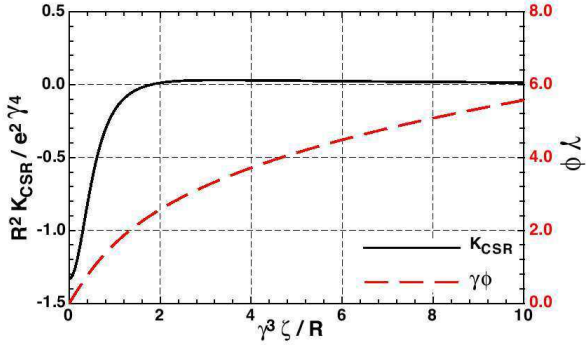}
\caption{$K_\Csr^\circ$ (left) and $\phi$ (right) as a function of $\zeta$
for a bend.\label{fg:kernel}}
\end{figure}

In Fig.~\ref{fg:2part}, $(v,w)$ are the coordinates of point $\BfP$ with
respect to point $\calO$ with the $v$--axis parallel to the orbit's
longitudinal $s$-axis at point $\calO$ and the $w$--axis
pointing upwards towards the inside of the element containing the point
$\BfP'$.  

With this notation, the difference in
$v$ and $w$ from the beginning of an element to the end is
\begin{align}
  \Delta v_i &= 
    \begin{cases}
      R_i \, ( \sin (\phi_i + \psi_i) - \sin \psi_i ) & \mbox{for a bend} \\
      d_i \, \cos \psi_i & \mbox{for a drift}
    \end{cases} \nonumber\\
  \Delta w_i &= 
    \begin{cases}
      R_i \, ( \cos \psi_i - \cos (\phi_i + \psi_i)) & \mbox{for a bend} \\
      d_i \, \sin \psi_i & \mbox{for a drift}
    \end{cases}
  \label{vrppp}
\end{align}
where $\psi_i$ is the orientation angle at the entrance end of the
element
\begin{equation}
  \psi_i = \sum_{k=1}^{i-1} \phi_{k}\ . 
\end{equation}
The above formulas are able to handle negative bends (beam rotating
clockwise). 
For a negative bend $R_i$, $g_i$ and $\phi_i$ are negative
while $d_i = R_i \, \phi_i$ is always positive.

With the assumption that all bend angles are small,
$v$ and $w$ can be approximated by
\begin{equation}
  v = \nu_1 - \nu_3 \, ,
  \qquad \mbox{and} \qquad
  w = \omega_2 \ ,
\end{equation}
where
\begin{align}
  \nu_1 &= \sum_{i=1}^{N} d_i
  \, , \qquad
  \omega_2 = \sum_{i=1}^{N}  d_i \left(
    \psi_i + \frac{1}{2} g_i \, d_i \right)\ ,
    \nonumber\\
  \nu_3 &= \sum_{i=1}^{N} d_i \left(
    \frac{1}{2} \psi_i^2 + \frac{1}{2} \psi_i \, g_i \, d_i +
    \frac{1}{6} g_i^2 \, d_i^2 \right)\ ,
    \label{eq:vrpnd}
\end{align}
and the small angles have been retained to second order.

The angle $\theta$ of the vector $\Bfn$ with respect to
the $v$--axis is $\theta = \sum_{i=1}^{N} g_i \, d_i$.
In terms of $v$ and $w$, the components of the vector $\BfL$ are
\begin{align}
  L_v &= v + R \, \sin \phi = [\nu_1 + d] - 
    \left[ \nu_3 + \frac{g^2 \, d^3}{6} \right]\ , \nonumber\\
  L_w &= w - R \, (1 - \cos \phi) = \omega_2 - \frac{g \, d^2}{2}\ , \\
  L &= \sqrt{L_v^2 + L_w^2} \nonumber\\
  &= [\nu_1 + d] - 
    \left[ \nu_3 + \frac{g^2 \, d^3}{6} -
    \frac{1}{8} \, \frac{(2 \, \omega_2 - g \, d^2)^2}{\nu_1 + d} \right]\ .
    \nonumber
\end{align}
Again angles are retained to second order. The path length is simply
\begin{equation}
  L_s = d + \sum_{i=1}^{N} d_i = d + \nu_1\ .
  \label{lrprpd}
\end{equation}
This, With Eq.~(\ref{eq:sslbl}), gives
\begin{equation}
  \zeta = \frac{\nu_1 + d}{2 \, \gamma^2} + 
    \left[ \nu_3 + \frac{g^2 \, d^3}{6} -  
    \frac{1}{8} \, \frac{(2 \, \omega_2 - g \, d^2)^2}{\nu_1 + d} \right]\ , 
    \label{eq:ssvrp2g}
\end{equation}
where terms to second order in combinations of angles and $1/\gamma$
are retained. Substituting these expressions into
Eq.~(\ref{eq:epegnb}), and defining
\begin{align}
  \alpha &= \gamma^2 \, \left(
    \omega_2 + g \, d \, \nu_1 + \frac{1}{2}\, g \, d^2 \right)\ , \\
  \kappa &= \gamma \, (\theta + g \, d)
  \, , \qquad
  \tau = \gamma \, (d + \nu_1)\ , \nonumber
\end{align}
the individual terms in Eq.~(\ref{eq:epegnb}) read as
\begin{align}
  \frac{1}{ (L - \BfL \cdot \beta \Bfn')^3} &= \frac{8 \gamma^9 \tau^3}{\left( \tau^2+\alpha^2 \right)^3} \ , \\
  \Bfn \cdot \left( \BfL - L \beta \Bfn'\right)/\gamma^2 &=   
  \frac{\tau^2 - \alpha^2 + 2 \, \tau \, \alpha \, \kappa}{2 \gamma^5 \tau}\ ,\nonumber \\
  \Bfn \cdot  \left( \BfL \times [(\BfL - L \beta \Bfn') \times \Bfa'] \right) /c^2 &= 
  \frac{g \, ( \tau^2 - \alpha^2 ) \, ( \alpha - \tau \, \kappa) }{2 \gamma^5 \tau}\ .\nonumber
\end{align}
Putting these together yields
\begin{align}
  K_\Csr \! = 4 \, r_c m c^2 \, \gamma^4 \, \tau^2 \, \Biggl\{
    &\frac{ g \, ( \tau^2 - \alpha^2 ) \, ( \alpha - \tau \, \kappa) }
         {\left( \tau^2 + \alpha^2 \right)^3} + {}
      \label{eq:kcsr_eq} \\
    & \qquad
    \frac{\tau^2 - \alpha^2 + 2 \, \tau \, \alpha \, \kappa}
         {\left( \tau^2 + \alpha^2 \right)^3}
      \Biggr\} -
    \frac{r_c m c^2}{\gamma^2 \, \zeta^2}\ . \nonumber
\end{align}
While we have used S.I. units, the classical radius $r_c$ and the mass
$m$ has been used to make the formula independent of the unit system.

From Eq.~(\ref{eq:kcsr_eq}), $K_\Csr^\circ$, which is $K_\Csr$  restricted
to the special case where points $\BfP$ and $\BfP'$ are within the
same bend, reduces to Eq.~(32) of Saldin \emph{et al.}~\cite{saldin97},
\begin{align}
  K_\Csr^\circ &=
  \frac{4 \, r_c m c^2 \, \gamma^4}{R^2}
  \left\{ \frac{\hat{\phi}^2/4 - 1}{2 \, (1 + \hat{\phi}^2/4)^3} + {} \right.
  \label{eq:k4egrt} \\
  & \left. \qquad
  \frac{1}{\hat{\phi}^2} \left[ \frac{1 + 3 \, \hat{\phi}^2/4}{(1 + \hat{\phi}^2/4)^3} -
  \frac{1}{(1 + \hat{\phi}^2/12)^2} \right] \right\}\ . \nonumber
\end{align}
where $\hat{\phi} \equiv \gamma \, \phi$, $\alpha = R \, \hat{\phi}^2
/2$, $\kappa = \hat{\phi}$, and $\tau = R \, \hat{\phi}$.  This
equation is valid for $\hat{\phi} > 0$; for $\hat{\phi} < 0$, $K_\Csr$
is, to a very good approximation, zero.

In the limit of small $\zeta$, $K_\Csr^\circ$ has a limiting value of
\begin{equation}
  K_\Csr^\circ (\zeta) \simeq \frac{-4  \, r_c m c^2 \, \gamma^4}{3 \, R^2}
  \qquad \mbox{for} \quad
  \zeta \ll \frac{R}{\gamma^3}\ .
\end{equation}
At large values of $\zeta$, $\zeta$ is cubic in $\phi$ so that $\phi \simeq
(24 \, \zeta /R)^{1/3}$. With this, Eq.~(\ref{eq:k4egrt}) becomes
\begin{equation}
  K_\Csr^\circ (\zeta) \simeq \frac{2  \, r_c m c^2}{3^{4/3} \, R^{2/3} \, \zeta^{4/3}}
  \qquad \mbox{for} \quad
  \zeta \gg \frac{R}{\gamma^3} \ ,
  \label{eq:k4e3rz}
\end{equation}
which corresponds to Eq.~(10) of Saldin \emph{et al.}~\cite{saldin97} (note the
error in the denominator of Saldin \emph{et al.}  Eq.~(10)).
Figure~\ref{fg:kernel} shows $K_\Csr^\circ (\zeta)$, which changes sign at
$\zeta \approx 1.8 \, R/\gamma^3$. The long tail at $\zeta > 1.8 \,
R/\gamma^3$ cannot be neglected since the integral $\int_0^\infty d\zeta
\, K_\Csr (\zeta)$ is zero. The vanishing of the integral is a
reflection of the fact that a closed loop of charged particles of
uniform density does not radiate.

\begin{figure}[tb]
\includegraphics[width=3.1in]{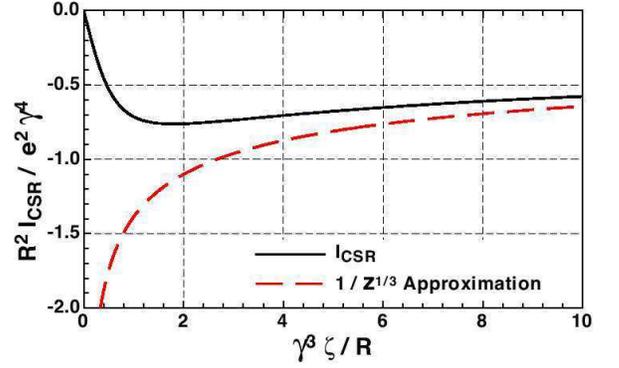}
\caption{$I_\Csr^\circ$ as a function of $\zeta$ for a bend.  The dashed
line is the large $\zeta$ approximation as given in
Eq.~(\ref{eq:iss2e3r}).\label{fg:integralK}}
\end{figure}

The fact that $K_\Csr$ is highly peaked in amplitude near $\zeta = 0$ can
be problematic for simulations at ultra--relativistic energies because
the characteristic longitudinal distance between particles or mesh
points needs to be less than $R/\gamma^3$.  One way of dealing with
the peaked nature of $K_\Csr$ is to first consider the kick from a
line of particles of density $\lambda(z)$ and then to integrate by
parts
\begin{align}
  \left(\frac{d{\cal E}}{ds}\right)_\Csr &=
    \int_{-\infty}^\infty \!\! dz' \, \lambda(z') \, K_\Csr(z-z')
    \label{eq:esslk} \\
  &= \int_{-\infty}^\infty \!\! dz' \,
  \frac{d\lambda(z')}{dz'} \, I_\Csr(z-z')\ ,
  \label{eq:esslsi}
\end{align}
where
\begin{align}
  I_\Csr(z-z') &= -\int_{-\infty}^{z'} \!\! dz'' \, K_\Csr(z-z'')
  \label{eq:isk}
\end{align}

$I_\Csr^\circ$, which is $I_\Csr$ for $\BfP$ and $\BfP'$ in the same
bend, is plotted in Fig.~\ref{fg:integralK}. The peaked nature of
$K_\Csr$ has been smoothed over at the cost of having to deal with a
derivative of $\lambda$. For $\zeta \gg R/\gamma^3$, the approximation of
Eq.~(\ref{eq:k4e3rz}) can be used to calculate an explicit
ultra-relativistic equation for $I_\Csr^\circ$ as in
\cite{stupakov02},
\begin{equation}
  I_\Csr^\circ(\zeta) = \frac{-2 \, r_c m c^2}{3^{1/3} \, R^{2/3}} \,
    \frac{1}{\zeta^{1/3}} \qquad \mbox{for} \quad \zeta \gg \frac{R}{\gamma^3}\ .
  \label{eq:iss2e3r}
\end{equation}
Equation~(\ref{eq:iss2e3r}) is also plotted in Fig.~\ref{fg:integralK}.

While, in general, it is helpful to have explicit formulas, for the
purposes of evaluation within a simulation program this is not
needed. The alternative is to use an exact implicit solution. Because
Eq.~(\ref{eq:ssvrp2g}) and Eq.~(\ref{eq:kcsr_eq}) are rational
functions, Eq.~(\ref{eq:isk}) can be integrated.  The last term that
compensates $K_\Sc$ can be integrated to $r_c m c^2 /(\gamma^2
\zeta)$. The other terms can be written with
\begin{equation}
\frac{\partial \zeta}{\partial d} = \frac{\tau^2+\alpha^2}{2 \gamma^2 \tau^2} 
\end{equation}
as
\begin{eqnarray}
  K_\Csr &=& 2 \, r_c m c^2 \gamma^2
    \left(\frac{\partial \zeta}{\partial d} \right)^{-1} \Biggl\{
    \frac{ g \, ( \tau^2 - \alpha^2 ) \, ( \alpha - \tau \, \kappa) }
         {\left( \tau^2 + \alpha^2 \right)^2}  {}
      \label{eq:kcsr_eq_v2} \nonumber\\
   &+& \frac{\tau^2 - \alpha^2 + 2 \, \tau \, \alpha \, \kappa}
         {\left( \tau^2 + \alpha^2 \right)^2}
      \Biggr\} 
    +\frac{\partial}{\partial \zeta} \left( \frac{r_c m c^2}{\gamma^2 \, \zeta}\right)\ .
\end{eqnarray}

With $\partial \tau / \partial d = \gamma$, $\partial \alpha /
\partial d = \gamma g \tau$, and $\partial \kappa / \partial d =
\gamma g$, one can further simplify to
\begin{align}
  K_\Csr \! = -2 \, r_c m c^2 \gamma \, &\left( 
    \frac{\partial \zeta}{\partial d} \right)^{-1} \frac{\partial}{\partial d} \left(
    \frac{ \tau + \alpha \kappa}{\tau^2 + \alpha^2 } \right)
      \label{eq:kcsr_eq_v3} \\
  &+ \frac{\partial}{\partial \zeta} \left( \frac{r_c m c^2}{\gamma^2 \, \zeta}\right)
  \ . \nonumber
\end{align}

This can be integrated over $\zeta$ to yield 
\begin{equation}
  I_\Csr(z, z') =-r_c m c^2\left(
  \frac{2 \, \gamma \, (\tau + \alpha \, \kappa)}
       {\tau^2 + \alpha^2} -
  \frac{1}{\gamma^2 \, \zeta}\right)\ .
  \label{eq:ieg2tt}
\end{equation}
It can be shown that, while quantities like $d$ and $g$
are discontinuous across element boundaries, $\tau$, $\alpha$, and
$\kappa$ are continuous and hence $I_\Csr$ is a continuous function as
it should be.

Equation~(\ref{eq:ieg2tt}) is the main result of this paper.  Using
Eq.~(\ref{eq:ieg2tt}), the integration of Eq.~(\ref{eq:esslsi}) in a
simulation program can be done via interpolation of
Eq.~(\ref{eq:ssvrp2g}).  Equation~(\ref{eq:ieg2tt}) has several
advantages over equations like Eq.~(\ref{eq:iss2e3r}). It is
applicable at lower values of $\gamma^3 \zeta$, that is, at lower
energies and/or smaller length scales. Additionally,
Eq.~(\ref{eq:ieg2tt}) has no singularity at small $\zeta$, and it can
be used to handle any combination of elements between the source and
kick points.

\begin{figure}[tb]
\includegraphics[width=3.1in]{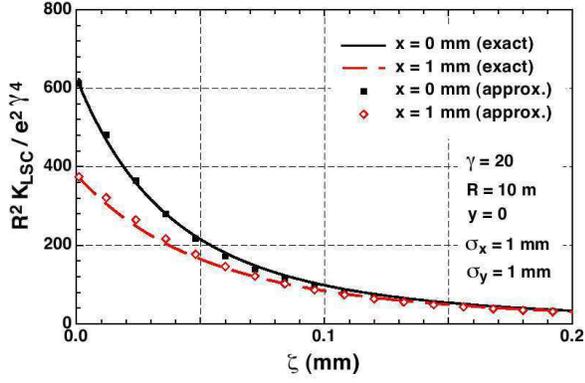}
\caption{Comparison between Eq.~(\ref{eq:kelsz}) and an exact
integration of Eq.~(\ref{eq:kls2p}).\label{fg:lsc_eq}}
\end{figure}

\section{Space Charge Calculation}
\label{k_sc}
The singularity at small $\zeta$ in the space charge term
$\BfE_\Sc$ is removed by considering the finite transverse beam
size. This term is equivalent to the problem of calculating the field
given a static distribution of charges. It will be assumed that at any
longitudinal position the transverse profile of the beam is
Gaussian. Thus, a longitudinal slice of the beam will produce
an energy change for a particle at longitudinal  $z$ and
transverse offset $(x, y)$ from the slice center of
\begin{align}
  d K_\Sc(x, y, z; z') \! &=
    \int_{-\infty}^{\infty} \int_{-\infty}^{\infty} \!\! dx' \, dy'
    \, \rho(x', y', z') \, dz'
    \label{eq:kls2p} \\
  & \qquad
    \frac{r_c m c^2 \, \gamma \, z}
    {(\gamma^2 \, (z-z')^2 + (x-x')^2 + (y-y')^2)^{3/2}}\ ,
    \nonumber
\end{align}
where $\rho$ is the bi-Gaussian distribution
\begin{equation}
  \rho(x, y, z) =
    \frac{\rho(z)}{2 \, \pi \, \sx \, \sy} \,
    \exp \left[
    -\frac{x^2}{2 \, \sx^2} - \frac{y^2}{2 \, \sy^2}
    \right]\ .
  \label{eq:rlpssx}
\end{equation}
A heuristic solution for Eq.~(\ref{eq:kls2p}) in the region of interest
($x \lesssim 3 \, \sx$ and $y \lesssim 3 \, \sy$) is
\begin{equation}
 d K_\Sc \approx \frac{ r_c m c^2 \, {\rm sign}(\zeta)\rho(z')dz'}
  {\sx \, \sy \, \exp
  \left[ \frac{x^2}{2 \, \sx^2} + \frac{y^2}{2 \, \sy^2} \right] +
  \frac{\sx^2 + \sy^2}{\sx + \sy} \, \gamma |\zeta| + \gamma^2\zeta^2}\ ,
  \label{eq:kelsz}
\end{equation}
where ${\rm sign}(z)$ is $1$ for positive and $-1$ for negative $z$,
and $\zeta = z-z'$.

Equation~(\ref{eq:kelsz}) is exact in the limit $z = 0$ and $z
\rightarrow \infty$, and is an excellent approximation in the region
in between. This is illustrated in Fig.~\ref{fg:lsc_eq}, which shows
$K_\Sc$ as a function of $z$ as computed from an integration of
Eq.~(\ref{eq:kls2p}) and from the approximate
Eq.~(\ref{eq:kelsz}). The particular parameters chosen for the
computation are given in the figure. Two cases were considered. One
where the kicked particle is on-axis, and the other where the kicked
particle is displaced by $\sx$ off-axis. As can be seen,
Eq.~(\ref{eq:kelsz}) gives an excellent approximation to the
longitudinal space charge kick.

At high energies, the CSR energy kick is independent of the beam energy
as indicated by Eq.~(\ref{eq:iss2e3r}). On the other hand, the factors
of $\gamma$ in the denominator of Eq.~(\ref{eq:kelsz}) insures the
$K_\Sc$ will be a decreasing function of $\gamma$.  Past some point in
an accelerator, the increasing $\gamma$ will make the effect of $K_\Sc$
small compared to the effect of $K_\Csr$. To estimate that point, consider
the maximum energy kick in a Gaussian bunch of $N_e$ particles with
\begin{equation}
  \rho(z) = \frac{N_e}{\sqrt{2 \, \pi} \, \sz} \, \exp
  \left[ - \frac{z^2}{2 \, \sz^2} \right] \ .
\end{equation}
The space charge kick is maximum at $x, y = 0$ and using
Eq.~(\ref{eq:kelsz}) gives
\begin{align}
  K_\Sc(0, 0, z) & \approx
    \frac{N_e  \, r_c m c^2}{\sqrt{2\, \pi} \, \sz} \label{eq:ke2ps} \\
  & \qquad
    \int_{0}^\infty \!\! d\zeta
    \frac{
    \exp \left[ -\frac{(z-\zeta)^2}{2 \, \sz^2} \right]
   - \exp \left[ -\frac{(z+\zeta)^2}{2 \, \sz^2} \right]}
    {\sx \, \sy + \frac{\sx^2 + \sy^2}{\sx + \sy} \, \gamma \, \zeta +
    \gamma^2 \, \zeta^2}\ .
    \nonumber
\end{align}
The dominant term in the denominator in the integrand is either $\sx
\sy$ for small $\zeta$, or $\gamma^2 \zeta^2$ for large $\zeta$. As an
approximation, the middle term in the denominator will therefore be
ignored.  The approximation $\sinh \left(z \zeta /\sz^2 \right)
\approx z \zeta/\sz^2$ will also be made. This approximation is
justified since, as shown below, the region of interest for $z$ is
around $z \approx \sz$ and for realistic beam parameters, $\sx\sy \ll
\gamma^2 \sz^2$ and therefore the only significant contribution to the
integral will come in the region $\zeta < \sz$. With these
approximations, Eq.~(\ref{eq:ke2ps}) becomes
\begin{equation}
  K_\Sc(z) \approx \frac{2\, N_e  \, r_c m c^2 \, z \,
  \exp \left[ -\frac{z^2}{2 \, \sz^2} \right]}
  {\sqrt{2\, \pi} \, \sz^3 }
  \int_{0}^\infty \!\! d\zeta \,
  \frac{\zeta \, \exp \left[ -\frac{\zeta^2}{2 \, \sz^2} \right]}
  {\sx \, \sy + \gamma^2 \, \zeta^2}\ .
\end{equation}
The integral can be evaluated to $(2\gamma^2)^{-1}e^a\Gamma(a)$ where
$\Gamma$ is the exponential integral and
$a=\sigma_x\sigma_z/(2\gamma^2\sigma_z^2)$. For small $a$ this is
approximately $-\ln(a)$.  The maximum $K_\Sc,\max$ occurs at $z =
\sz$. Using $\sx \sy \ll \gamma^2 \sz^2$ gives
\begin{align}
  K_\Sc,\max &\approx -\frac{\exp \left[ -\frac{1}{2} \right]}{\sqrt{2 \, \pi}} \,
    \frac{N_e \, r_c \, m \, c^2}{\sz^2 \, \gamma^2} \,
    \log \left( \frac{\sx \, \sy}{2 \, \gamma^2 \, \sz^2} \right) \nonumber\\
  &\approx -\frac{N_e \, r_c \, m \, c^2}{4 \, \sz^2 \, \gamma^2} \,
    \log \left( \frac{\sx \, \sy}{2 \, \gamma^2 \, \sz^2} \right)\ .
\end{align}
The maximum CSR kick at large energies is, from
Eq.~(\ref{eq:kcsr_eq}),
\begin{equation}
  K_\Csr,\max \approx 0.8 \, \frac{N_e \, r_c \, m \, c^2}{(R^2 \, \sz^4)^{1/3}}\ .
\end{equation}
The ratio is
\begin{equation}
  \frac{K_\Sc,\max}{K_\Csr,\max} \approx
  \left[ 0.3 \,
  \log \left( \frac{2 \, \gamma^2 \, \sz^2}{\sx \, \sy} \right)
  \right] \,
  \frac{1}{\gamma^2} \,
  \left( \frac{R}{\sz} \right)^{2/3}\ .
  \label{eq:kk2gs}
\end{equation}
The condition for $K_\Sc$ begin small can be written as
\begin{equation}
  \gamma \gg
  \left[ 0.3 \,
  \log \left( \frac{2 \, \gamma^2 \, \sz^2}{\sx \, \sy} \right)
  \right]^{1/2} \,
  \left( \frac{R}{\sz} \right)^{1/3}\ .
  \label{eq:g2gss}
\end{equation}
The condition for this can be well approximated by noting that the
quantity in square brackets on the LHS of Eq.~(\ref{eq:g2gss}) is
slowly varying and never extremely large.  Thus the required condition
is
\begin{equation}
  \gamma \gg
  M \left( \frac{R}{\sz} \right)^{1/3}\ ,
\end{equation}
where $M$ is a number of order unity. In this case, $\sz$ must be
interpreted as the characteristic longitudinal distance over which the
bunch density is changing.

\section{CSR in Bmad}

\begin{figure}[tb]
\includegraphics[width=3.1in]{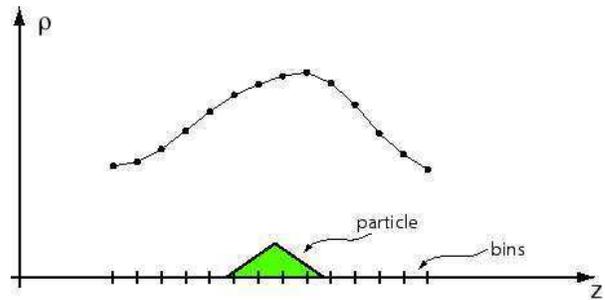} \caption{{\tt Bmad}
implementation of the CSR algorithm. The beam of particles is divided
up into a number of bins. The contribution of a particle to a bin's
total charge is determined by the overlap of the particle's triangular
charge distribution and the bin.\label{fg:bin}}
\end{figure}

The above algorithm for simulating CSR and Longitudinal Space Charge
(SC) has been implemented as part of the {\tt Bmad}~\cite{b:bmad}
subroutine library for relativistic charged-particle simulations. {\tt
Bmad} simulates a beam as a set of particles.  The beam is tracked
through a lattice element by dividing the element into a number of
slices. Tracking through a slice involves first propagating the
particles independently from each other and then applying the CSR and
SC energy kicks.  To calculate the energy kick, the beam is divided
longitudinally into $N_b$ bins as shown in Fig.~\ref{fg:bin}. For
computing the charge in each bin, each beam particle is considered to
have a triangular charge distribution. The overlap of the triangular
charge distribution with a bin determines that particle's contribution
to the total charge in that bin. The width of the particle's
triangular charge distribution and the number of bins are set by the
user. The bin width is dynamically adjusted at each time step so that
the bins will span the bunch length.  Increasing the particle width
smooths the distribution at the cost of resolution.

The charge density $\lambda_i$ at the center of the $i^{th}$ bin is
taken to be $\lambda_i = \rho_i / \Delta z_b$ where $\rho_i$ is the total
charge within the bin and $\Delta z_b$ is the bin width. The charge density
is assumed to vary linearly in between the bin centers. The CSR energy
kick for a particle at the center of the $j^{th}$ bin after traveling a
distance $ds_\mb{slice}$ according to Eq.~(\ref{eq:esslsi}) is then
\begin{equation}
  d\calE_j = ds_\mb{slice} \,
  \sum_{i = 1}^{N_b} \,
  \left( \lambda_i - \lambda_{i-1} \right) \,
  \frac{I_\Csr(j-i) + I_\Csr(j-i+1)}{2}\ , 
  \label{eq:dedsl}
\end{equation}
where
\begin{equation}
  I_\Csr(j) \equiv I_\Csr(z = j\, \Delta z_b)\ .
\end{equation}

Evaluation of $I_{\Csr, j}$ involves inversion of
Eq.~(\ref{eq:ssvrp2g}) to obtain $d$.  Because $z$ is a monotonic
function of $d$, Newton's method~\cite{b:nr} is used to find numbers
$d_1$ and $d_2$ which bracket the root and then Ridders'
Method~\cite{b:nr} is used to quickly find $d$.

In deriving Eq.~(\ref{eq:dedsl}), the approximation
\begin{equation}
  \int_{j \, \Delta z_b}^{(j+1) \, \Delta z_b} \! dz \, I_\Csr(z) \approx
  \Delta z_b \, \frac{I_\Csr(j) + I_\Csr(j+1)}{2}
\end{equation}
has been used. Generally this is an excellent approximation, except
when $j = 0$ and $\Delta z_b \gg R / \gamma^3$, as shown in
Fig.~\ref{fg:integralK}. Here, however, the integral can be done
exactly assuming that the source and kick points lie within the same
element
\begin{equation}
  \int_{0}^{\Delta z_b} \! dz \, I_\Csr(z) =
  \frac{1}{\gamma^2} \, \ln \left( \frac{2 \, \gamma^2 \, \Delta z_b}{d(\Delta z_b)} \right) -
  \frac{d(\Delta z_b)^2 \, g^2}{4}\ .
\end{equation}

Once the energy kick at the centers of the bins is calculated, the
energy kick applied to a particle is calculated via interpolation
assuming a linear variation of the kick between bin centers. 

In calculating the energy kick, The computational time for calculating
the charge in the bins charge scales as $N_p$, the number of particles
in the simulation. The computational time for calculating the energy
kick at the bin centers scales as $N_b^2$, and the time for
calculating the energy kick of the particles scales as $N_p$.

\subsection{Chamber Walls}

The simulation incorporates the shielding of the top and bottom
chamber walls by using image currents. Appendix \ref{sc:shielding}
explains why neglecting the width of a chamber is a good approximation
when it's width is larger than its hight. Here, because the image
current is well separated from the actual beam, there are no
singularities to deal with, $K_\Sc$ does not have to be subtracted,
and a straight forward integration is done using Eq.~(\ref{eq:esslk}),
and Eq.~(\ref{eq:kcsr_eq}).
\begin{align}
  d\calE_j&(\mb{image}) = 2 * ds_\mb{slice} \\
 &\times \sum_{k = 1}^{N_i} \, (-1)^k 
  \sum_{i = 1}^{N_b} \,
  q_i * K(z=(j-i)\Delta z_b, y = k \, h)\ ,\nonumber
\end{align}
where $q_i = \lambda_i \, \Delta z_b$ is the charge in a bin, $h$ is
the chamber height, and $k$ indexes the image currents at vertical
displacement $y = \pm k \, h$.  The number of image charges $N_i$
needs to be chosen large enough so that the neglected image currents
do not have a significant effect on the simulation results.  Because
the relevant angles are not small, the image charge kick $K$ must be
calculated without the small angle approximation, as in
Eq.~(\ref{eq:kcsr_eq}).

\section{Agoh and Yokoya CSR Calculation}

Agoh and Yokoya (A\&Y) have developed a code to calculate CSR wake fields by
directly integrating Maxwell's equations on a mesh representing a
rectangular beam chamber~\cite{agoh04}. The approach depends on the
paraxial approximation, a rigid Gaussian bunch density, small chamber
dimensions relative to the bending radius, and ultra-relativistic
particles. Using a co-moving coordinate system $(x,y,s)$ in Fourier
space, they are able to reduce the problem to a tractable
two-dimensional differential equation,
\begin{equation}
  \frac{\partial}{\partial s}\textbf{E}_\perp =
  \frac{i}{2 k}\left[\left( \nabla^2_\perp +
  \frac{2 k^2 x}{R}\right)
  \textbf{E}_\perp - \frac{1}{\epsilon_0}\nabla_\perp \rho_0 \right]\ ,
\end{equation}
where $\textbf{E}_\perp$ is a complex 2-dimensional vector related to
the perpendicular electric field, $\rho_0$ is the charge density, $k$
is the wave number, and $R$ is the magnet bending radius. It is solved
using a finite-differencing method. 

\section{CSR in elegant}

The particle tracking code {\tt elegant} (version 17.2.2) uses
Eq.~(\ref{eq:ssyfree}) to compute CSR kicks without shielding by a vacuum
chamber \cite{borland01}. The charge distribution
$\lambda(z)$ and its derivative $d\lambda/dz$ are calculated by
binning the macroparticles and then employing a smoothing filter.

\section{CSR Wake Formula}

\subsection{Transient Effects at Magnet Edges}

Using retarded fields, Saldin \emph{et al.}~\cite{saldin97} derive, in the
ultra-relativistic limit, a formula for the wake-field due to a bunch
entering from a drift region into a bend
\begin{eqnarray}
  \left( \frac{d{\cal E}}{ds} \right) &=&
  -\frac{2 N_e r_c m c^2}{3^{1/3} R^{2/3}} \left\{
  \frac{\lambda(s-s_L)-\lambda(s-4 s_L)}{s_L^{1/3}} \right.
  \label{eq:ssyfree}\\
  &+& \left. \int_{s-s_L}^{s}
  \frac{1}{(s-s')^{1/3}}\frac{d\lambda}{d s'}d s' \right\}\ ,\ \ \nonumber
  s_L \equiv \frac{R \phi^3}{24}\ ,
\end{eqnarray}
where $\phi$ is the angle traveled into the magnet by the bunch
center. Eq.~(\ref{eq:ssyfree}) reduces to the free space steady-state
wake-field of Eq.~(\ref{eq:iss2e3r}) in the limit $s_L \rightarrow \infty$.

As worked out by Emma and Stupakov \cite{stupakov02}, synchrotron
radiation will continue to propagate and affect the bunch beyond the
end of a bending magnet. For a finite magnet of length $L_m$, an
ultra-relativistic bunch at a distance $x$ from the end of this magnet
experiences the free space exit wake-field
\begin{align}
  \left( \frac{d{\cal E}}{ds} \right)_{\textrm{exit}}&=
  N_e r_c m c^2 \left\{
  4\frac{\lambda(s-\Delta s(L_m))}{L_m +2 x} \right.
  \label{eq:esexit}\\
  & \left.  - \int_{0}^{L_m}
  \frac{4}{l+2x}\lambda' (s-\Delta s(l)) 
  \frac{\partial \Delta s(l)}{\partial l}d l \right\}, \nonumber \\
  \Delta s(l) &\equiv \frac{l^3}{24 R^2} \frac{l+4 x}{l+x}\ .
\end{align}

\subsection{Steady State CSR in a Bend}

CSR effects in a vacuum chamber have been computed by the Green's
function of grounded parallel plates~\cite{schwinger45,
warnock90}. These formulas are difficult to compute numerically, due
to the presence of high order Bessel functions, so we will use an
excellent approximation developed by A\&Y
\cite{agoh04}. The impedance for the steady-state in a dipole with
horizontal plates separated by a distance $h$ is
\begin{align}
  Z(k) &= Z_0 \frac{2 \pi}{h}\left( \frac{2}{k R} \right)^{1/3}
    \sum_{p=0}^{\infty} F_{AY}(\beta_p^2)\ ,\\
  F_{AY}(x) &\equiv \textrm{Ai}'(x)(\textrm{Ai}'(x)-i \, \textrm{Bi}'(x))\\
  & +x \, \textrm{Ai}(x)(\textrm{Ai}(x)-i\, \textrm{Bi}(x))\ , \nonumber \\
  \beta_p &\equiv (2p+1)\frac{\pi}{h}\left(\frac{R}{2 k^2}\right)^{1/3}\ ,
\end{align}
where $Z_0 = c \mu_0$ is the free space impedance, $k$ is the wave
number, and Ai and Bi are Airy functions. The parallel plate
wake-field due to a bunch with longitudinal density $\lambda(z)$ is
obtained by Fourier transform:
\begin{align}
  \left( \frac{d{\cal E}}{ds} \right)_\textrm{pp} &=
  -N_e r_c m c^2~  \Re \left(
  \frac{1}{\pi} \int_0^\infty Z(k) \tilde \lambda(k) e^{i k s} dk \right)\ ,
  \label{eq:PPagoh}\\
  \tilde \lambda(k) &= \int_{-\infty}^{\infty} \lambda(z) e^{-i k z} dz\ .
\end{align}

\section{Comparison between Bmad, the Agoh and Yokoya, and elegant}

\begin{table}
\centering
\caption{Parameters used in simulations}

\begin{tabular*}{\columnwidth}{@{\extracolsep{\fill}}lcccccc}
\hline
Set & $R$(m) & $L$(m) & $w$(cm) & $h$(cm) &  $\sigma_z$(mm)  & $E_0(keV/m/e)$\\
\hline
A         & 10.0	&	3.0	&	50.0	&	2.0	&	0.3	&	2.48  \\
B         & 10.0	&	1.0	&	34.0	&	28.0	&	0.3	&	2.48 \\
C         & 10.0	&	1.0	&	10.0	&	10.0	&	0.3	&	2.48 \\
D         & 10.0	&	1.0	&	60.0	&	10.0	&	0.3	&	2.48  \\
E         & 1.20	&	0.419	&0.762	&	0.3	&	0.036	&	3704.1\\
F         & 2.22	&	0.678	&	1.71	&	2.54	&	1.0	&	29.22\\
G        & 87.9	&	6.574	&	8.0	&	4.0	&	0.3	& 12.52\\
\hline
\end{tabular*}

\label{tab:params}
\end{table}

In order to validate our method, we compare simulations from {\tt
Bmad} to those using the A\&Y code and {\tt elegant}. For
ease of reading, all magnet and bunch parameters used are enumerated
in Tab.~\ref{tab:params}, and will be referred to by a letter.

Note that all electric fields in the graphs are normalized by
\begin{equation}
  E_0 = \frac{2 N_e r_c m c^2}{\sqrt{2\pi}(3 R^2 \sz^4)^{1/3}}\ ,
\label{eq:e0}
\end{equation}
which approximately describes the maximum amplitude of the CSR-Wake
$d{\cal E} / d s $.

\begin{figure}[t]
\centering
\includegraphics[width=\columnwidth]{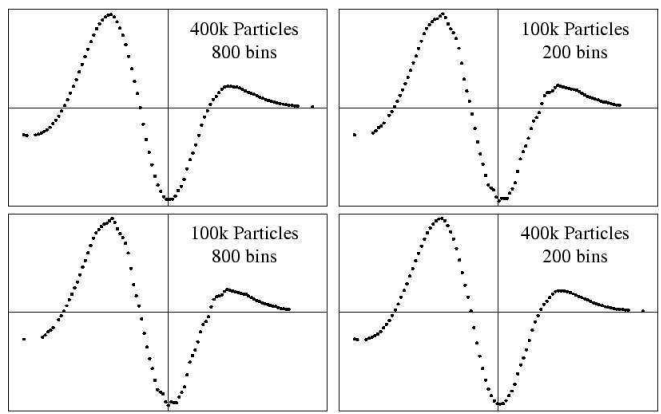}
\caption{The number of particles and the number of bins are varied in
{\tt Bmad} using parameter set A.\label{fg:parameters}}
\end{figure}

All simulations use a bunch charge of 1nC, and an energy of 1 GeV,
unless otherwise noted. We used {\tt Bmad} with the following
parameters: number of bins $N_b=800$, number of macro-particles $N_p =
4\cdot 10^5$, number of image charges $N_i=32$, and tracking step
$ds_\mb{step}=1$mm. The triangular bin width, as in Fig.~\ref{fg:bin},
is 32 bins. Figure~\ref{fg:parameters} shows that this choice of $N_b$
and $N_p$ is reasonable by varying $N_p$ and $N_b$, using parameter
set A. While reducing the number of particles leads to a visibly less
smooth wake field. The number of bins has been increased, until the
wakefield starts becoming less smooth. The other parameters were
similarly varied to find the applied parameters.

\begin{figure*}[ht!]
\centering
\includegraphics[width=6in]{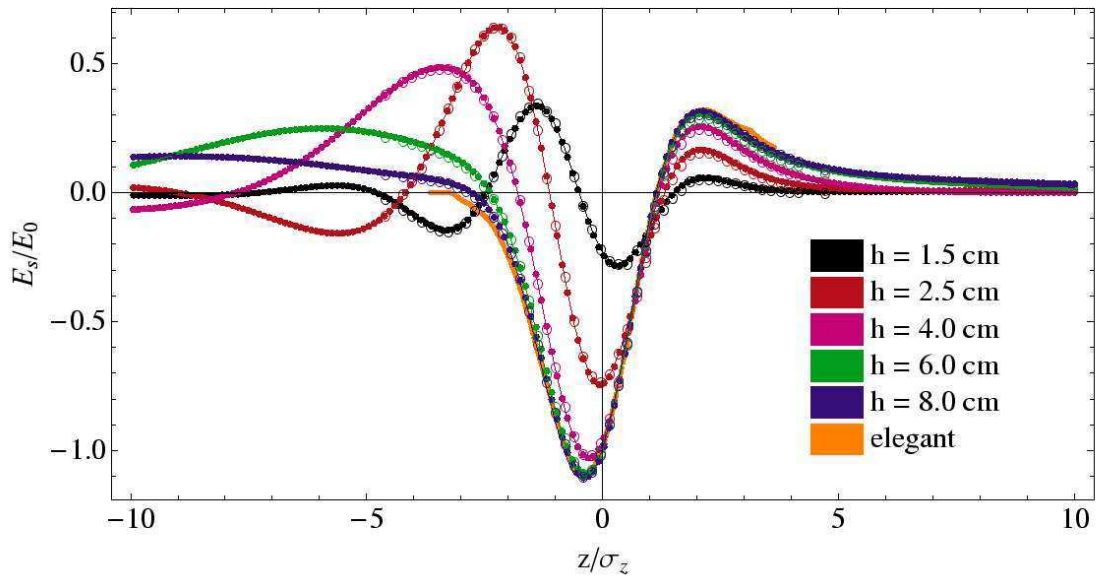}
\caption{The steady state using parameter set A for varying chamber
heights $h$. The A\&Y code (dots), {\tt Bmad} (circles), and the
CSR-Wake formula Eq.~(\ref{eq:PPagoh}) (lines) agree well, and {\tt
elegant} agrees with the data at large chamber
heights.\label{fg:agohBmadmesh}}
\end{figure*}

\subsection{Steady State Case}

Figure~\ref{fg:agohBmadmesh} shows the steady-state CSR kick in a bend
as a function of $z$ for various values of the chamber height.  The
parameters used correspond to set A of Table~\ref{tab:params} and are
the same as used for Fig.~1 of A\&Y \cite{agoh04}.

Figure~\ref{fg:agohBmadmesh} shows excellent agreement among the {\tt
CSRmesh} code, {\tt Bmad}, and the CSR-Wake formula
Eq.~(\ref{eq:PPagoh}). Note that {\tt Bmad} computes the wake by
tracking a bunch, and therefore shows the result only in the range of
the length of the bunch. This is not problematic, because the wake
only influences particles in the region of the bunch.

\subsection{Transient Case}

\begin{figure*}[t!]
\centering
\begin{tabular}{cc}
\includegraphics[width=\columnwidth]{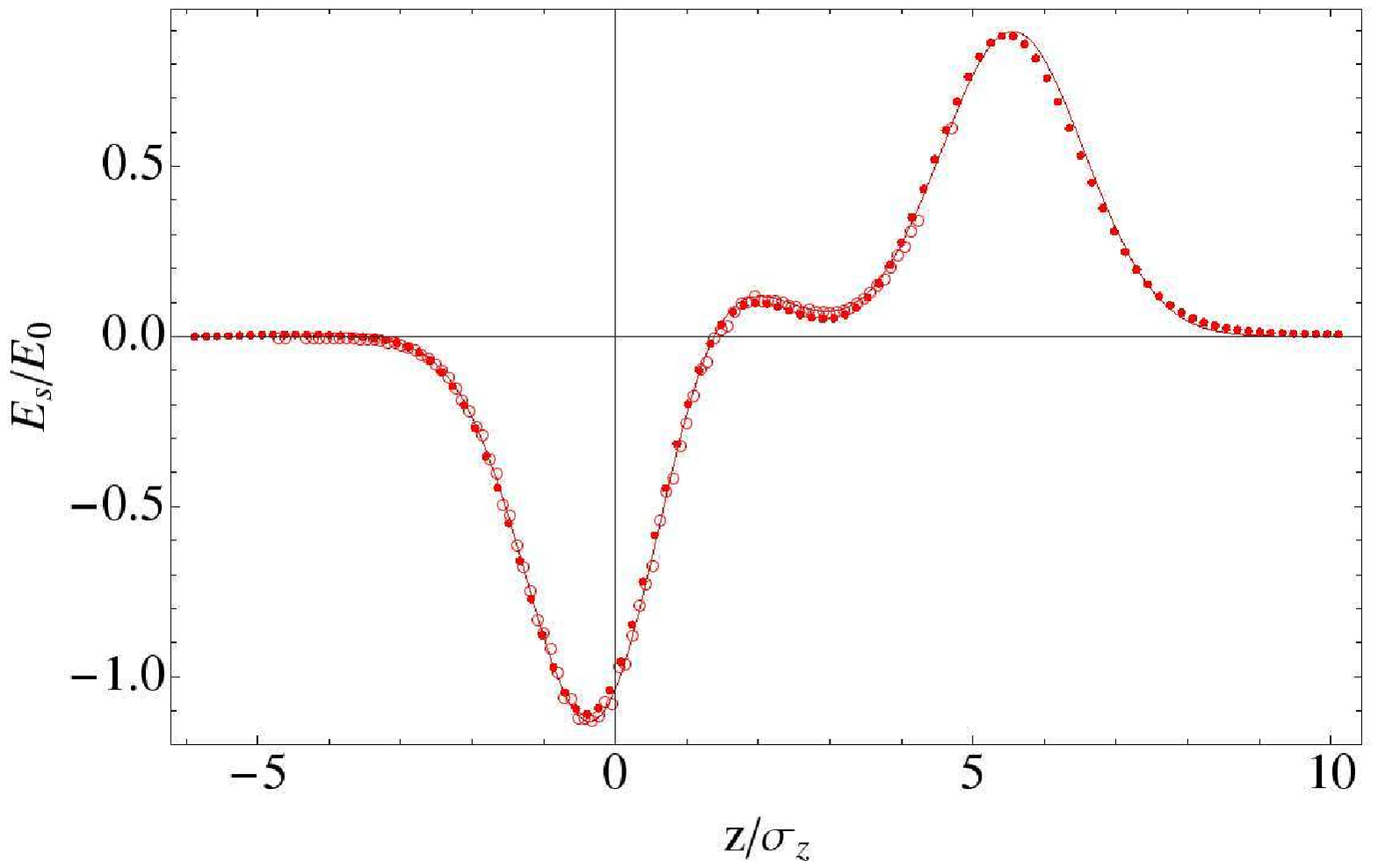} &
\includegraphics[width=\columnwidth]{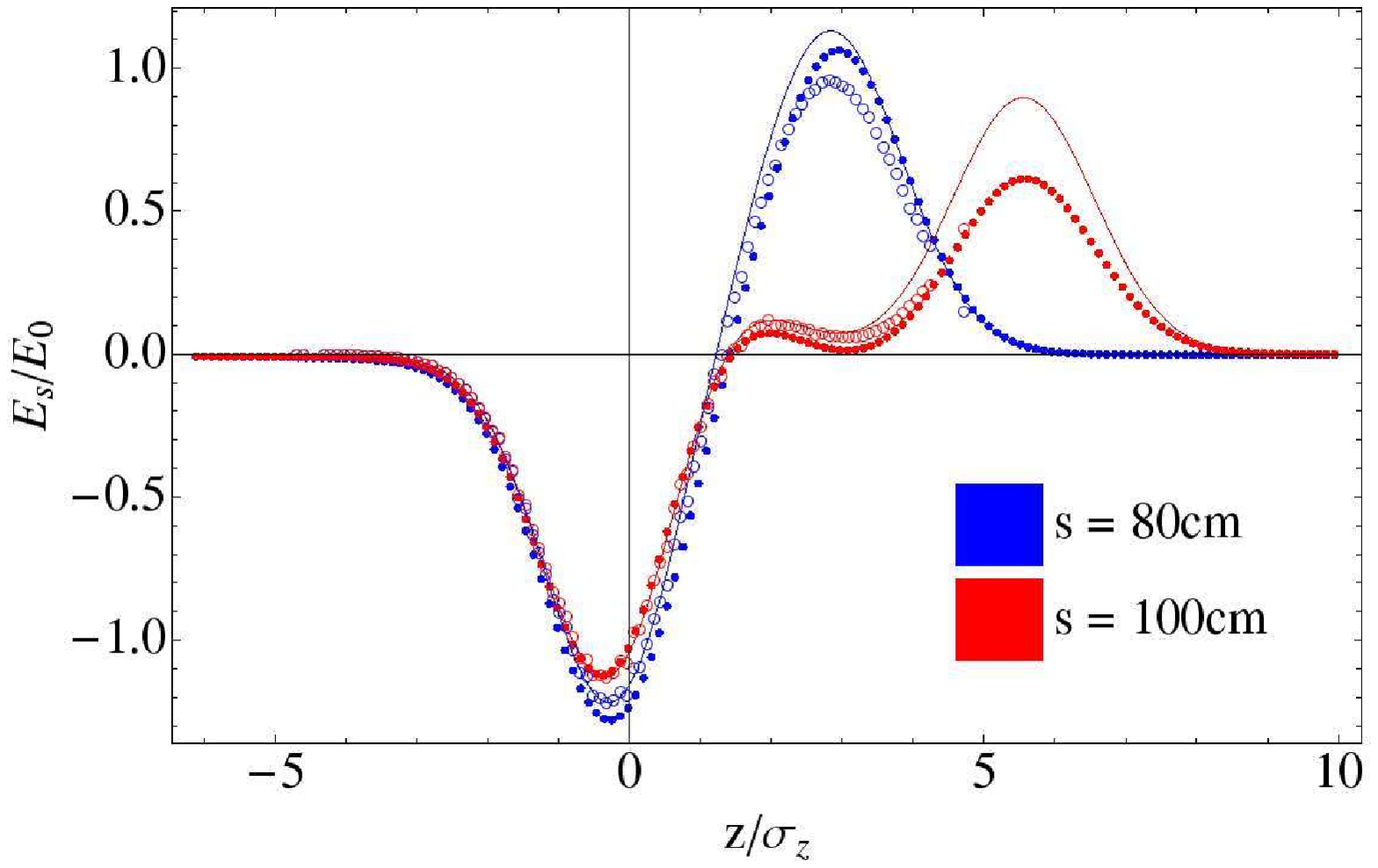}
\end{tabular}
\caption{The transient case for a large chamber using parameter set B
with a large chamber (left), and set C with a smaller chamber
(right). The A\&Y code (dots), {\tt Bmad} (circles), and the
CSR-Wake formula Eq.~(\ref{eq:ssyfree}) (lines) agree well in the
former case, but differ in the latter case.\label{fg:setBC}}
\end{figure*}

\begin{figure}[t]
\centering
\includegraphics[width=\columnwidth]{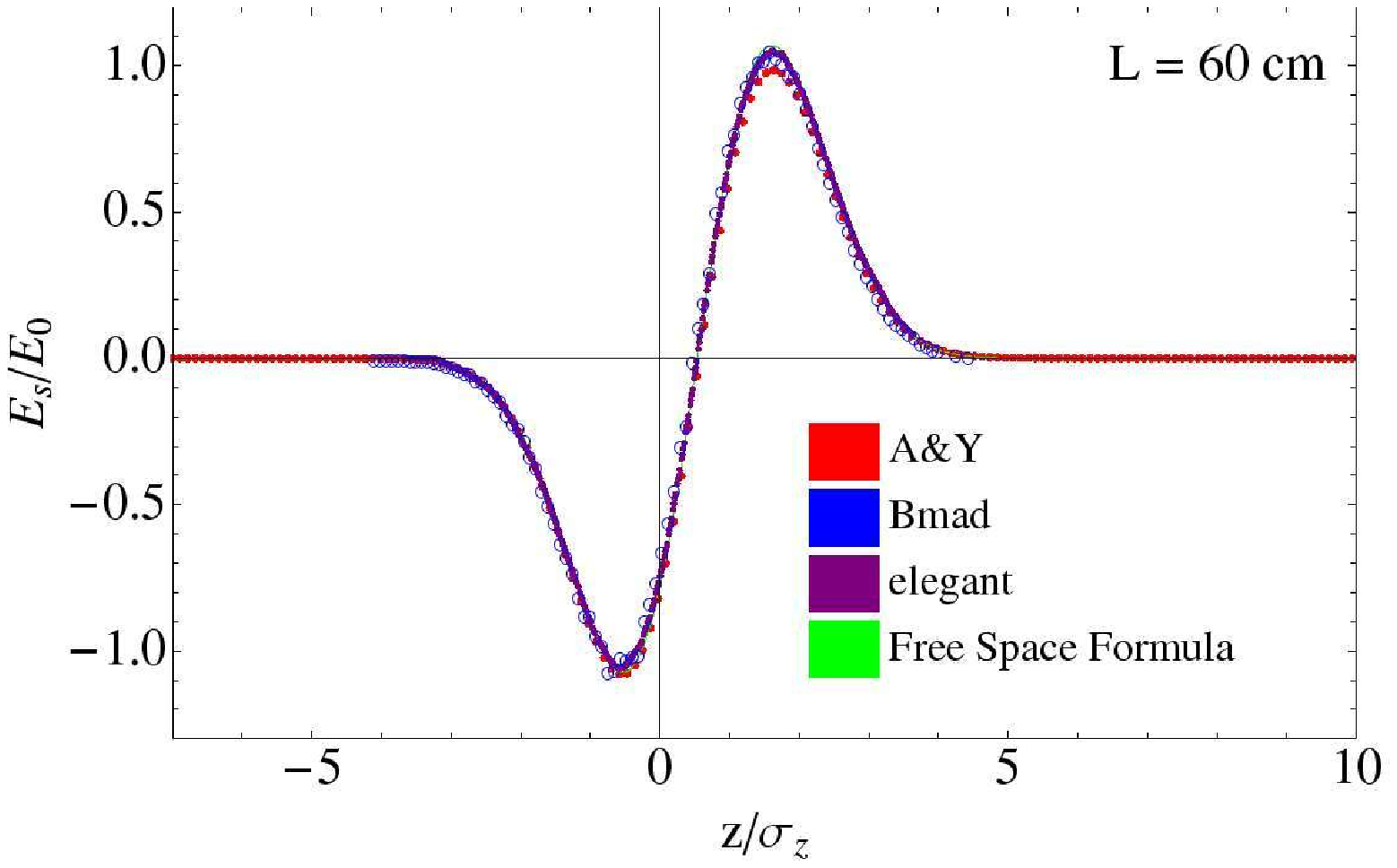}
\includegraphics[width=\columnwidth]{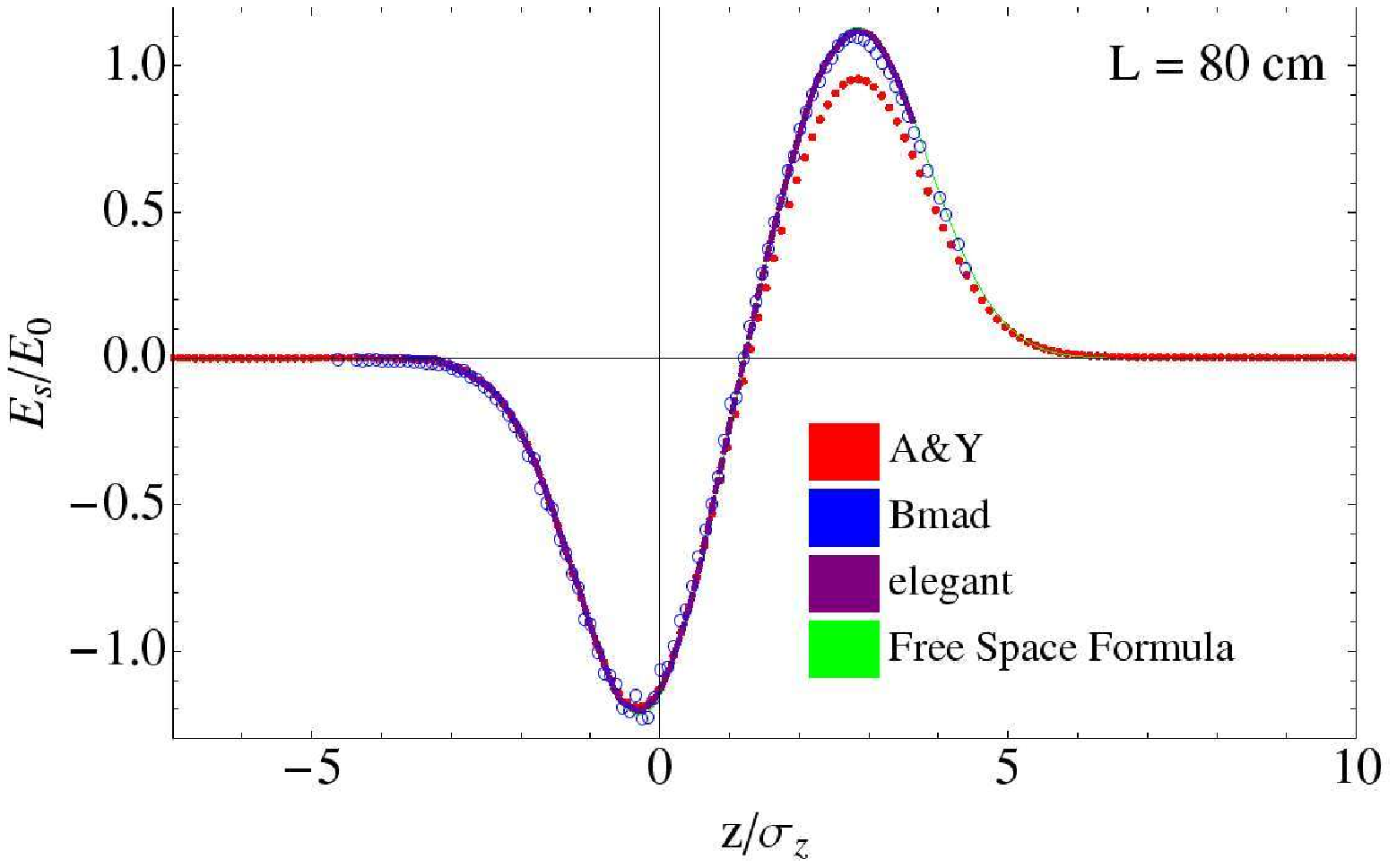}
\includegraphics[width=\columnwidth]{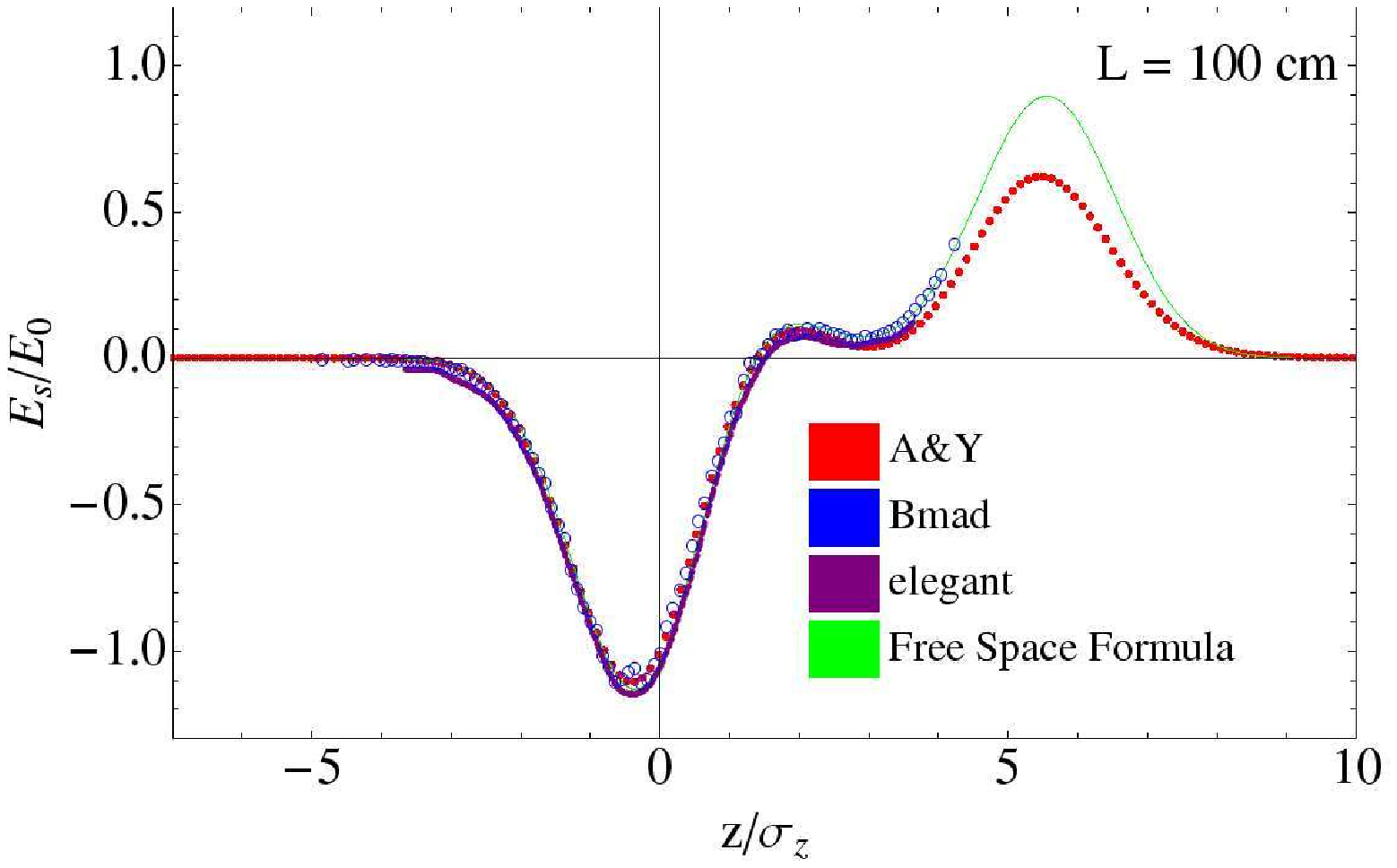}
\caption{Length dependence for the transient cases using the wide
chamber in parameter set D.  Magnet lengths of 60cm (top), 80cm
(middle), 100cm (bottom) show that the codes deviate for longer
magnets, but {\tt Bmad} (circles) agrees with CSR-Wake
Eq.~(\ref{eq:ssyfree}). Thus, reduction of the chamber height and
longer magnet length lead to discrepancies of codes.\label{fg:setD}}
\end{figure}

\begin{figure}[tb]
\centering
\includegraphics[width=\columnwidth]{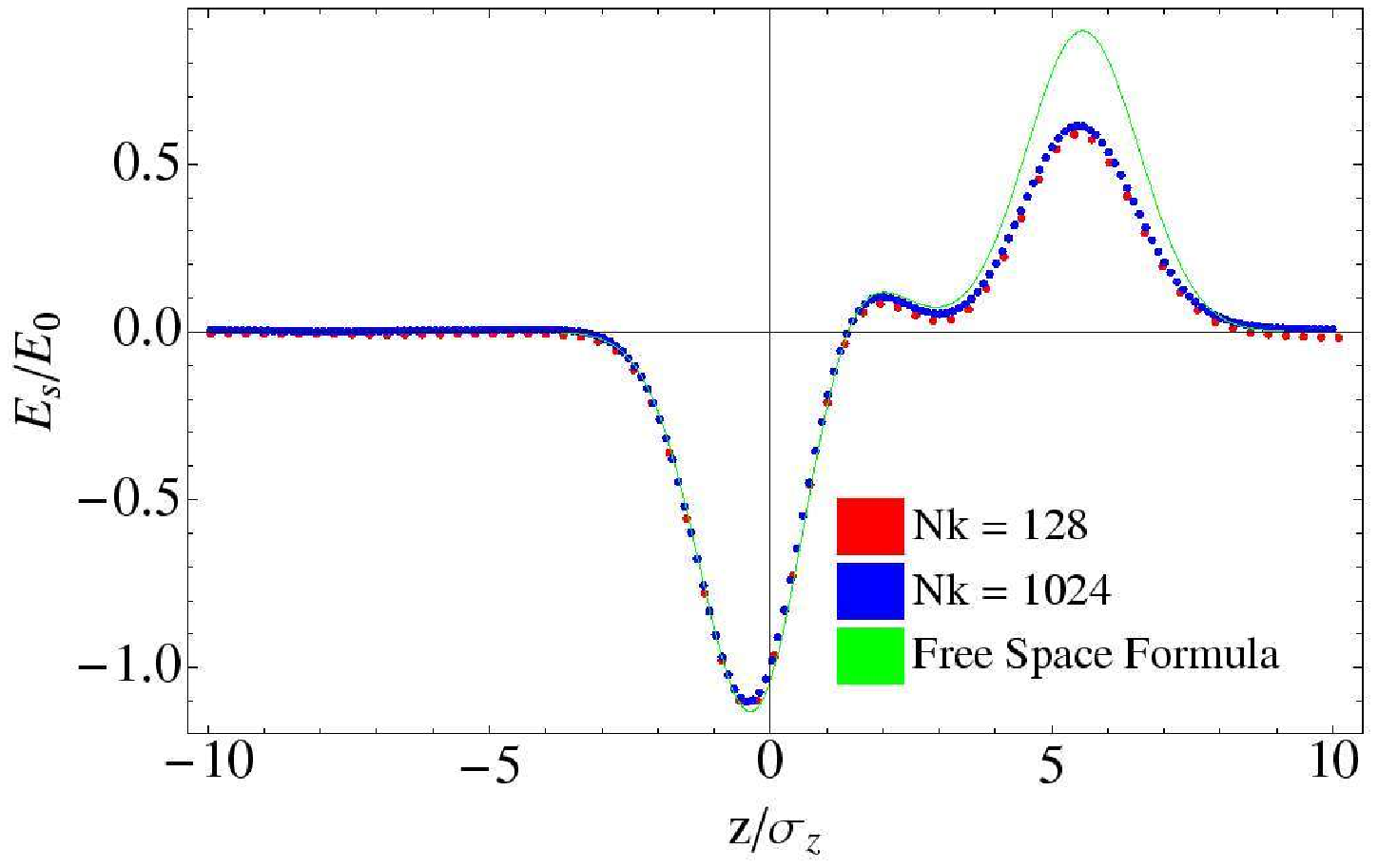}
\caption{Using parameter set D, varying the number of Fourier
coefficients (Nk) used in the A\&Y calculation does not
change the numerical solution of the Maxwell equations, and therefore
does not account for deviations in the presented comparison of
codes.\label{fg:nk}}
\end{figure}

Parameter sets B, C, and D are used to explore the transient case
where some of the kick is generated from particles in the drift region
before the bend.  Set B corresponds to values used in Fig.~3 of Agoh
and Yokoya \cite{agoh04}, which has a chamber of relatively large
cross section. Figure~\ref{fg:setBC} (left) shows agreement between
{\tt Bmad}, the A\&Y code, and the CSR-Wake formula
Eq.~(\ref{eq:ssyfree}). However, for parameter set C in
Fig.~\ref{fg:setBC} (right), where the chamber is smaller as in Fig.~4
of \cite{agoh04}, a discrepancy appears in the front of the
bunch. This discrepancy is noted in \cite{agoh04} and explained to be
due to the finite chamber size.

\begin{figure}[t]
\centering
\includegraphics[width=\columnwidth]{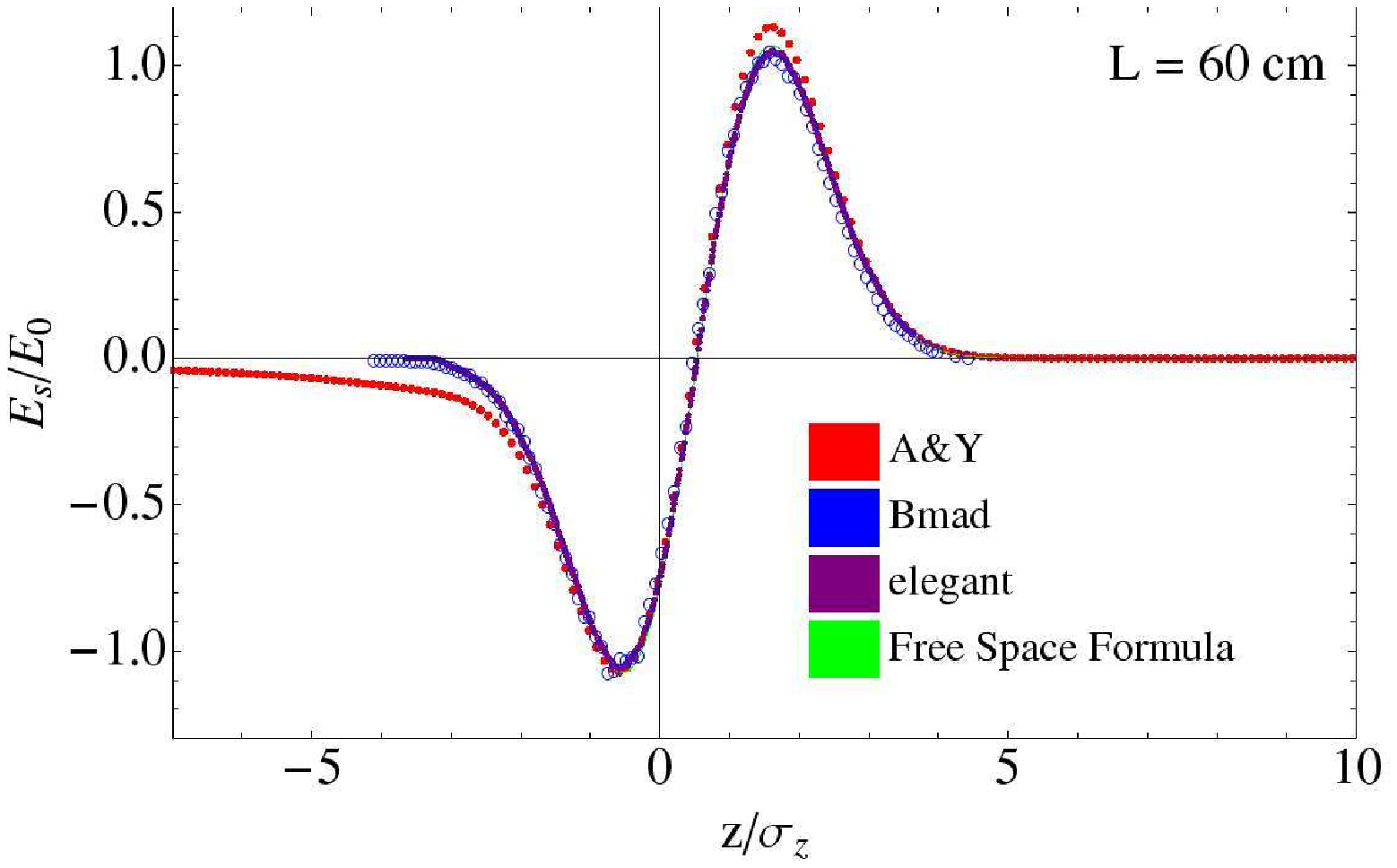}
\includegraphics[width=\columnwidth]{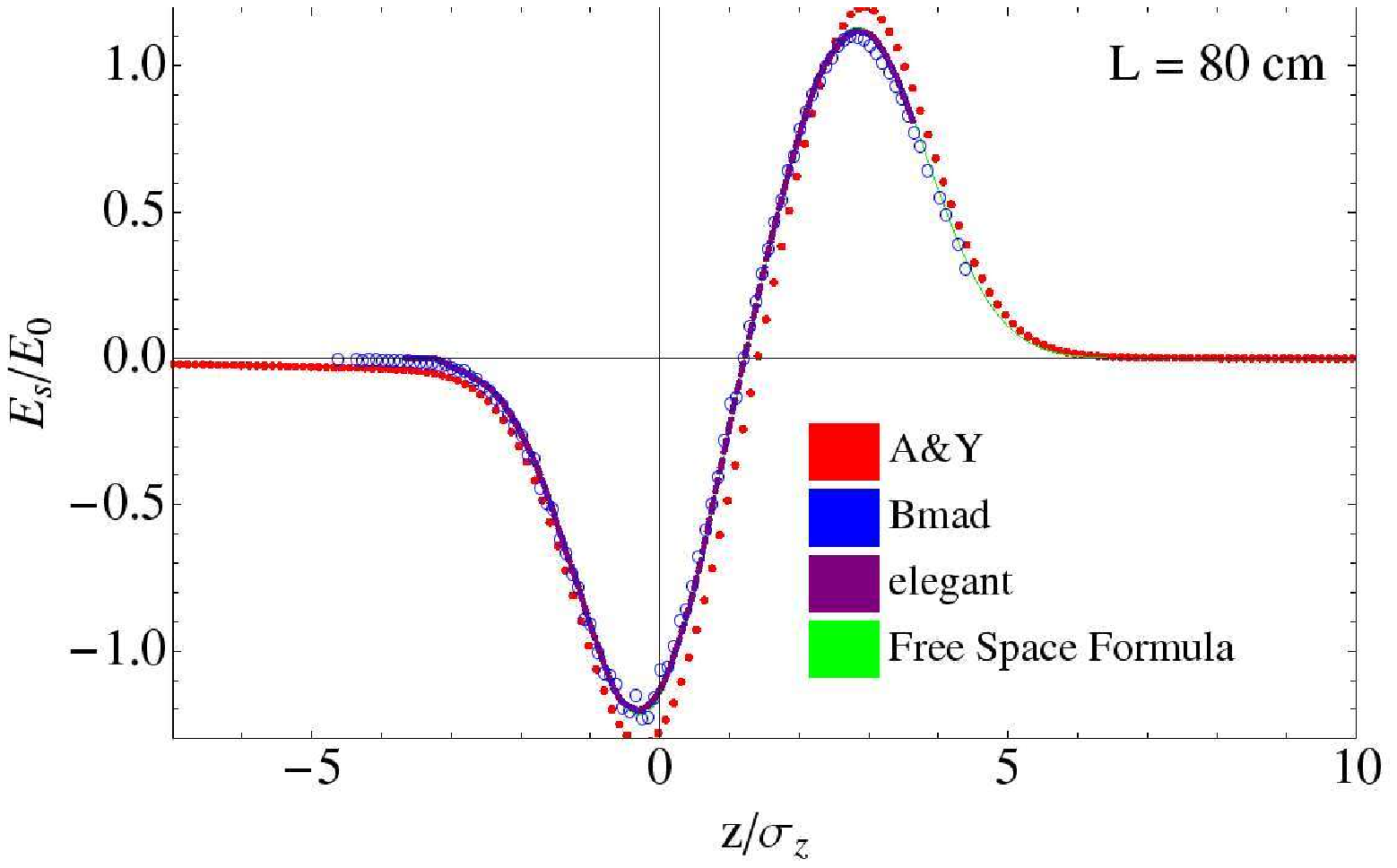}
\includegraphics[width=\columnwidth]{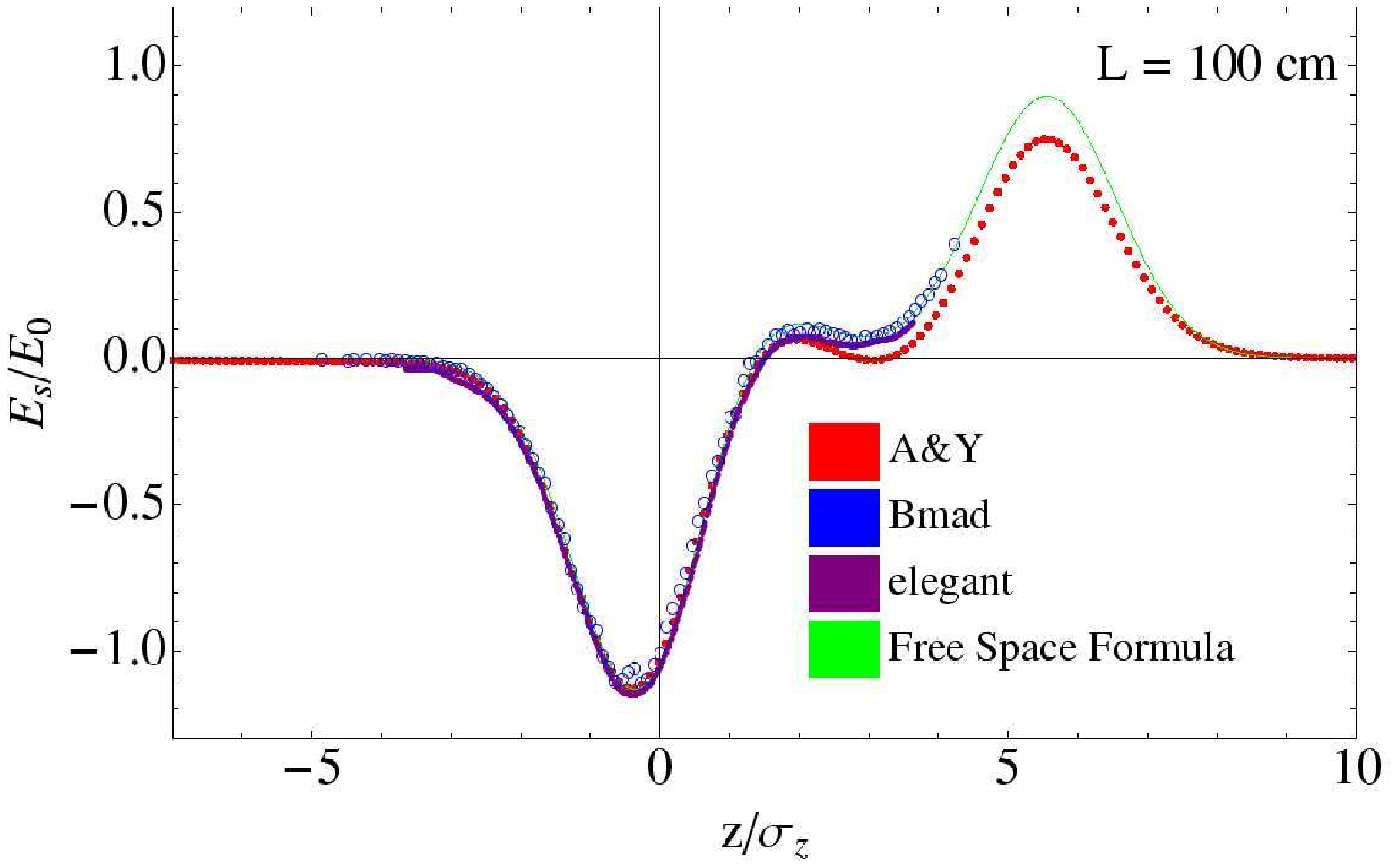}
\caption{Length dependence for the transient cases using a tall
chamber by reversing width and height in parameter set D.  Magnet
lengths are 60cm (top), 80cm (middle), and 100cm
(bottom).\label{fg:setDRev}}
\end{figure}

\begin{figure}[t]
\centering
\includegraphics[width=\columnwidth]{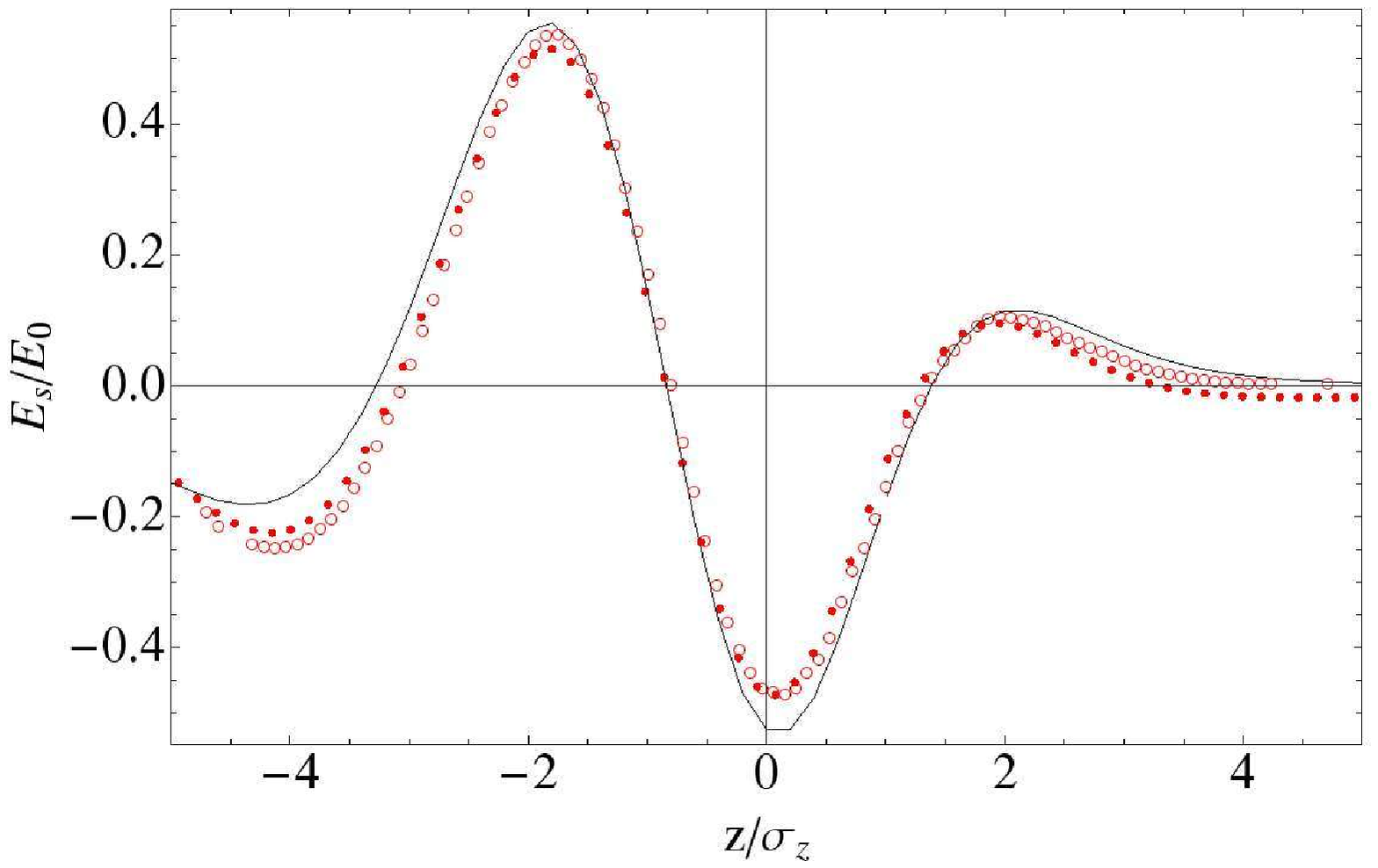}
\includegraphics[width=\columnwidth]{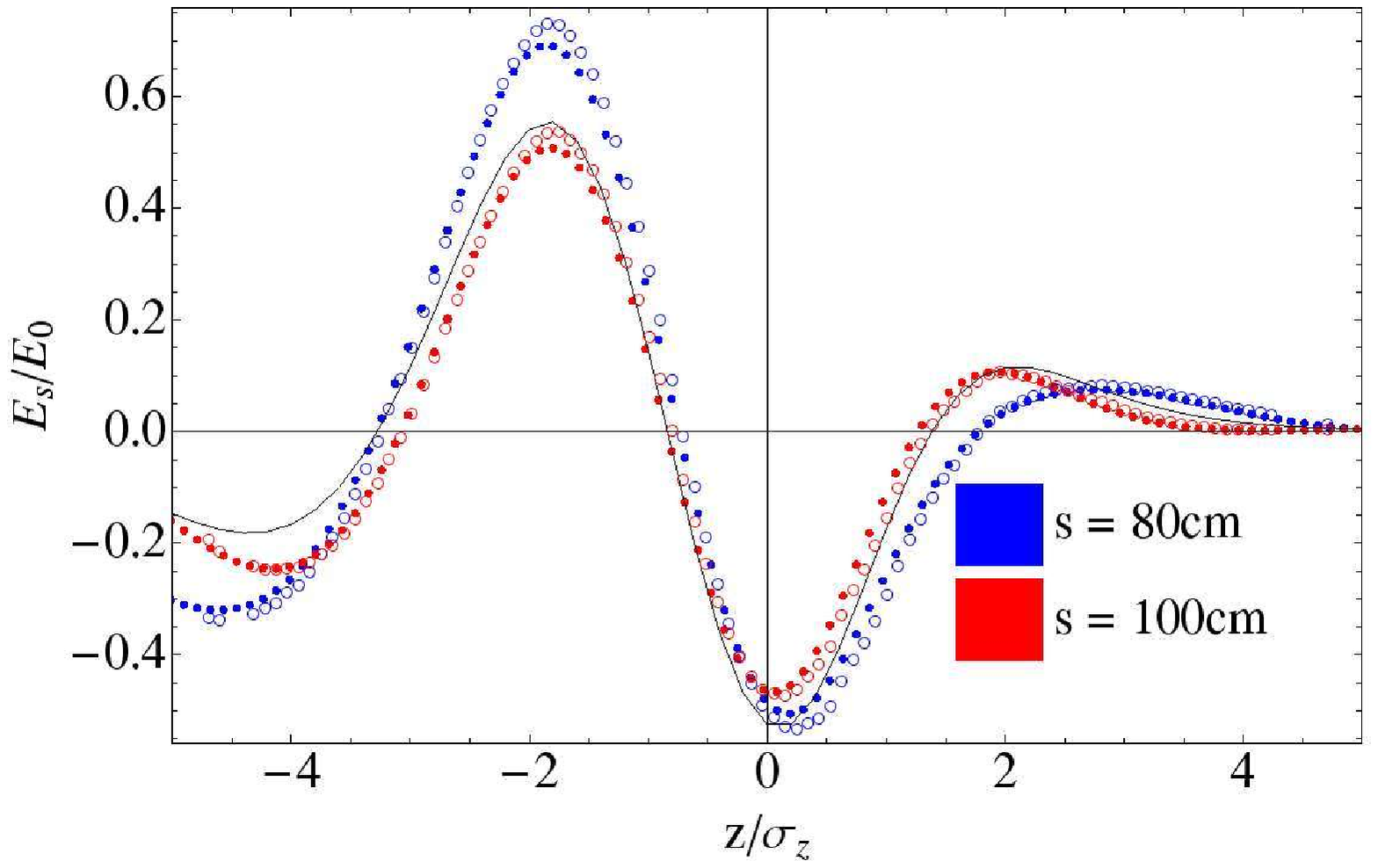}
\caption{The same as Fig.~\ref{fg:setBC}, except with the chamber
height reduced to 2cm. A\&Y code (dots), {\tt Bmad}
(circles), and the shielded steady-state CSR-Wake
Eq.~(\ref{eq:PPagoh}) (line) show slight deviations from one
another.\label{fg:setBC_shielded}}
\end{figure}

We use parameter set D to analyze whether the chamber width is
responsible for this difference, and one sees in Fig.~\ref{fg:setD}
that the effect remains and thus appears to be due to the reduced
chamber height. It also is larger for longer magnets. Varying the
number of Fourier coefficients used in the A\&Y calculation
in Fig.~\ref{fg:nk} does not change this result either, verifying that
we chose a reasonable number of Fourier coefficients for solving the
Maxwell equations numerically. Reversing the set D height and width,
as in Fig.~\ref{fg:setDRev}, somewhat reduces the discrepancy. Thus,
again indicating that the effect is apparently due to the reduced
chamber height. Practically speaking, the difference in these
wake-fields only becomes appreciable far in front of the bunch, where
there are few particles to affect. To see how strong the deviation
becomes for especially small chamber heights, we repeat parameter sets
B and C with $h = 2$cm in Fig.~\ref{fg:setBC_shielded}, and the
disagreement between {\tt Bmad} and the A\&Y code is again
not very large but significant.

\subsection{Realistic Magnets}

\begin{figure}[ht!]
\centering
\includegraphics[width=\columnwidth]{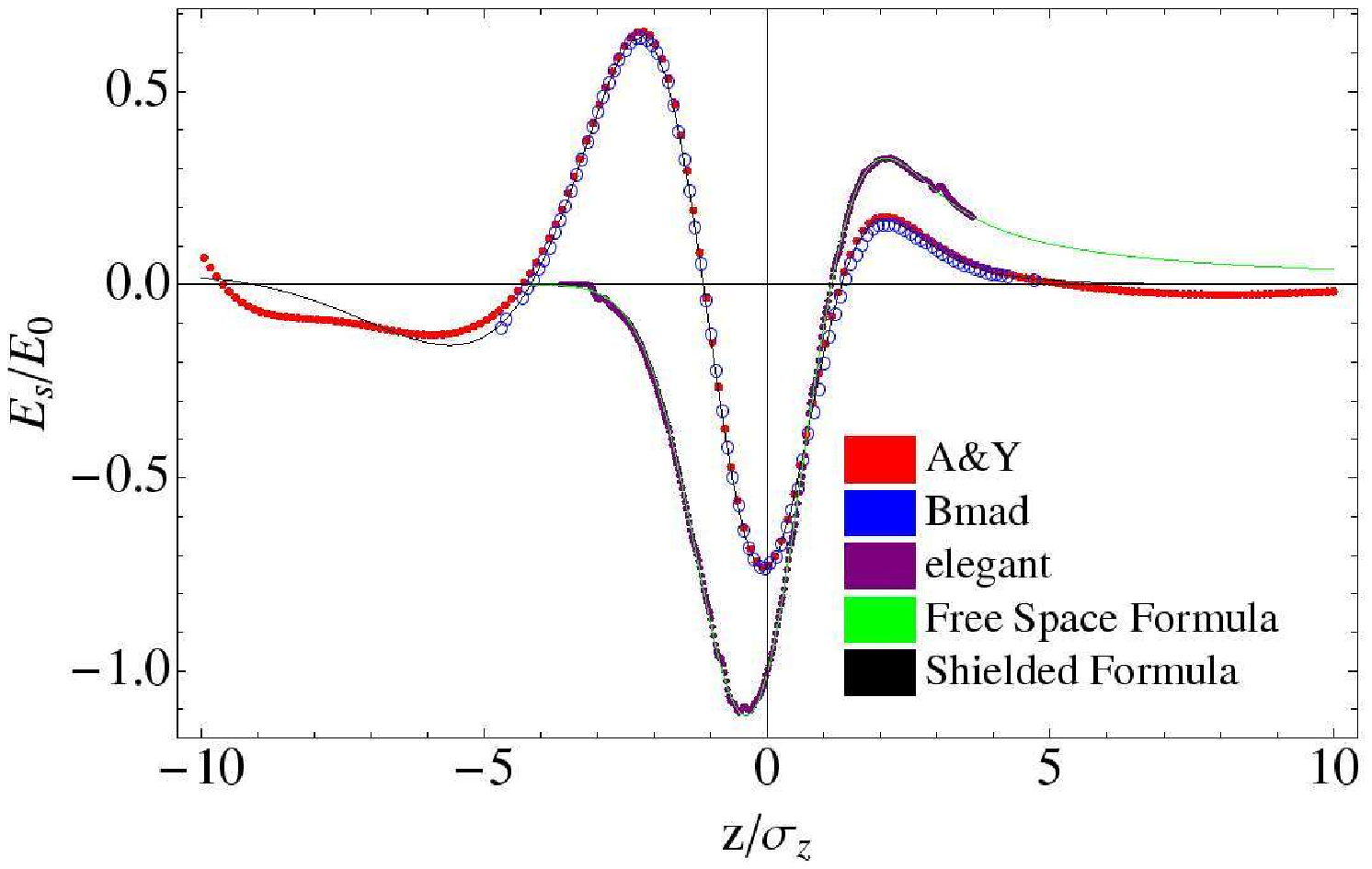}
\includegraphics[width=\columnwidth]{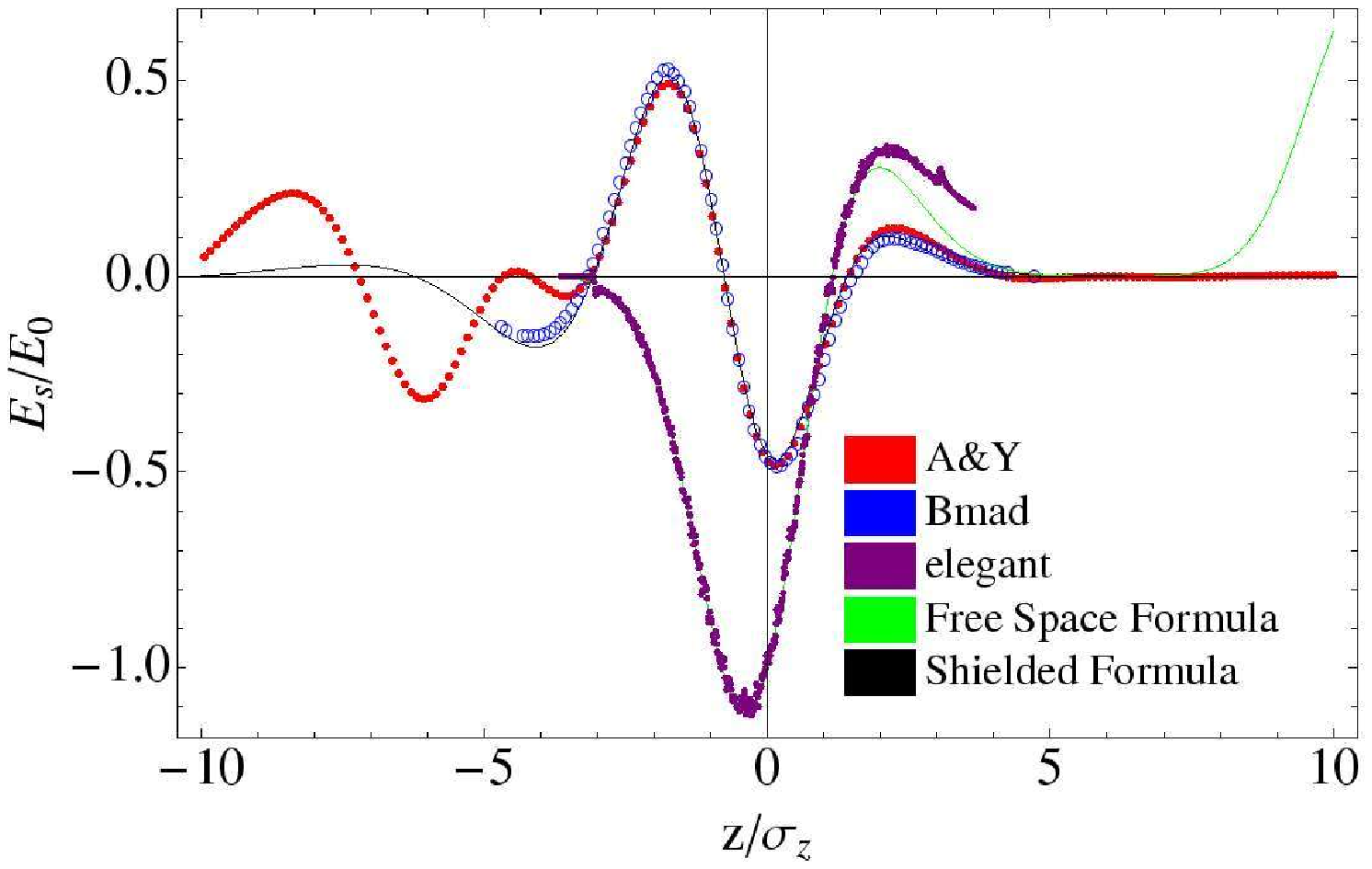}
\caption{Realistic magnets: Parameter set E (JLab TH2 magnet) line
(top), set F (CESR Analyzer magnet) (bottom). {\tt Bmad} agrees with
the CSR-Wake formula Eq.~(\ref{eq:PPagoh}) better than the other codes
at the bunch tail.\label{fg:setEF}}
\end{figure}

\begin{figure}[t]
\centering
\includegraphics[width=\columnwidth]{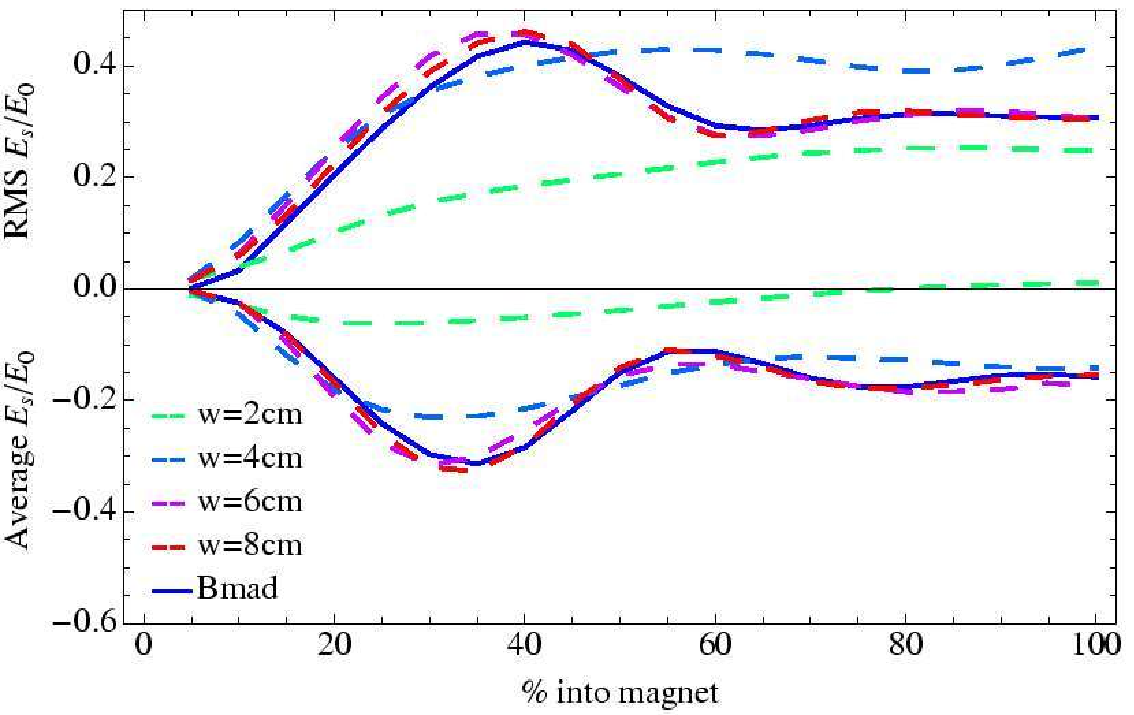}
\caption{Average and RMS longitudinal electric fields for parameter
set G (the Cornell ERL's CE magnet) for various chamber widths using the
A\&Y code, compared to {\tt Bmad}, which has infinite chamber
width.\label{fg:setG}}
\end{figure}

To evaluate how significant the differences are in realistic magnets,
we use parameter sets E, F, and G which correspond to the JLab TH2, CESR
Analyzer magnet, and Cornell ERL CE magnets, respectively. Wake-fields for
the first two are plotted in Fig.~\ref{fg:setEF}, showing good
agreement between the A\&Y code and {\tt Bmad}. The free
space steady-state wake-field Eq.~(\ref{eq:ssyfree}) and results from
{\tt elegant} are plotted for reference.

The principal detrimental effects of the CSR-Wake in an accelerator
are energy loss and increase in energy spread of a bunch. The
transverse bunch distribution can also be damaged and is mostly
influenced by increases of the energy spread which, through dispersive
orbits, couples to transverse motion. To visualize CSR-driven energy
loss and energy spread in a single magnet, we plot in
Fig.~\ref{fg:setG} the average and RMS electric field across the bunch
distribution as a function of distance into the magnet. This is done
using parameter set G, where we examine how well one can ignore the
chamber width, as done in the {\tt Bmad} calculation. One sees that,
in this shielded case, the chamber width $w$ in the A\&Y code
begins to change the wake field when it is comparable or less than
4cm, the height of the chamber. This effect is heuristically explained
in appendix \ref{sc:shielding}. It again shows that ignoring the
chamber width, as in {\tt Bmad} is a reasonable approximation when the
chamber is wider than high.

It is apparent that shielding by a vacuum chamber reduces the power
emitted by CSR very effectively. It does not reduce the energy spread
nearly as much, and is therefore not as effective for preserving bunch
properties as one might have concluded from the reduced radiation
power. Interestingly, the 4cm wide chamber even produces \emph{larger}
RMS $E_s$ than wider chambers.

\subsection{Exit Wake}

\begin{figure}[t]
\centering
\includegraphics[width=\columnwidth]{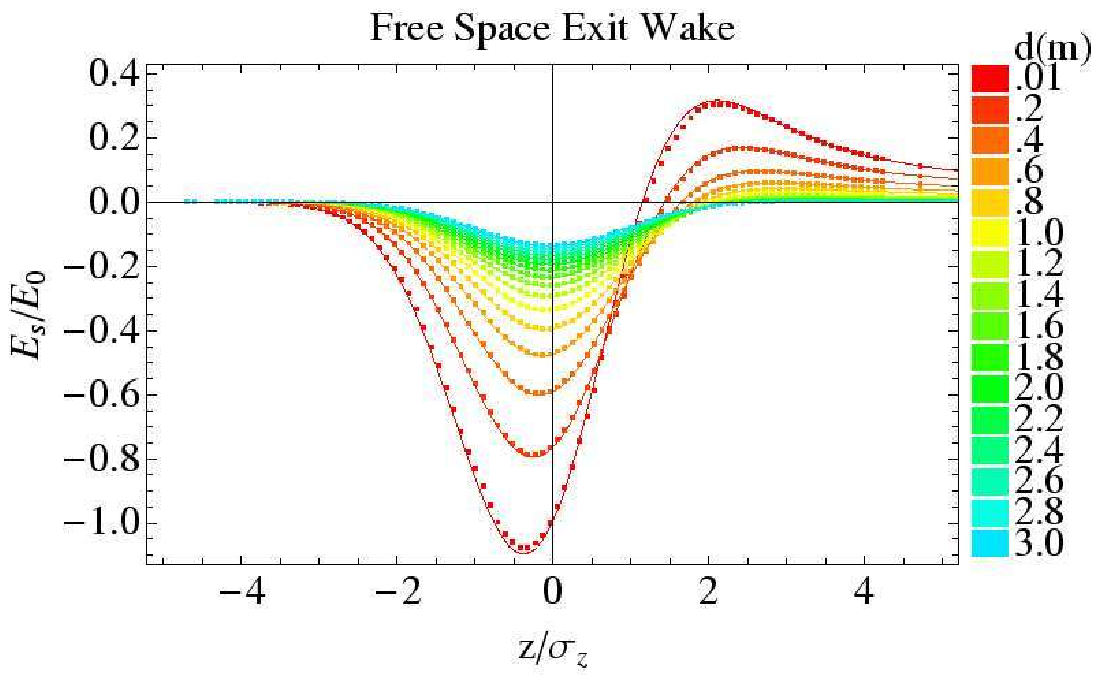}
\includegraphics[width=\columnwidth]{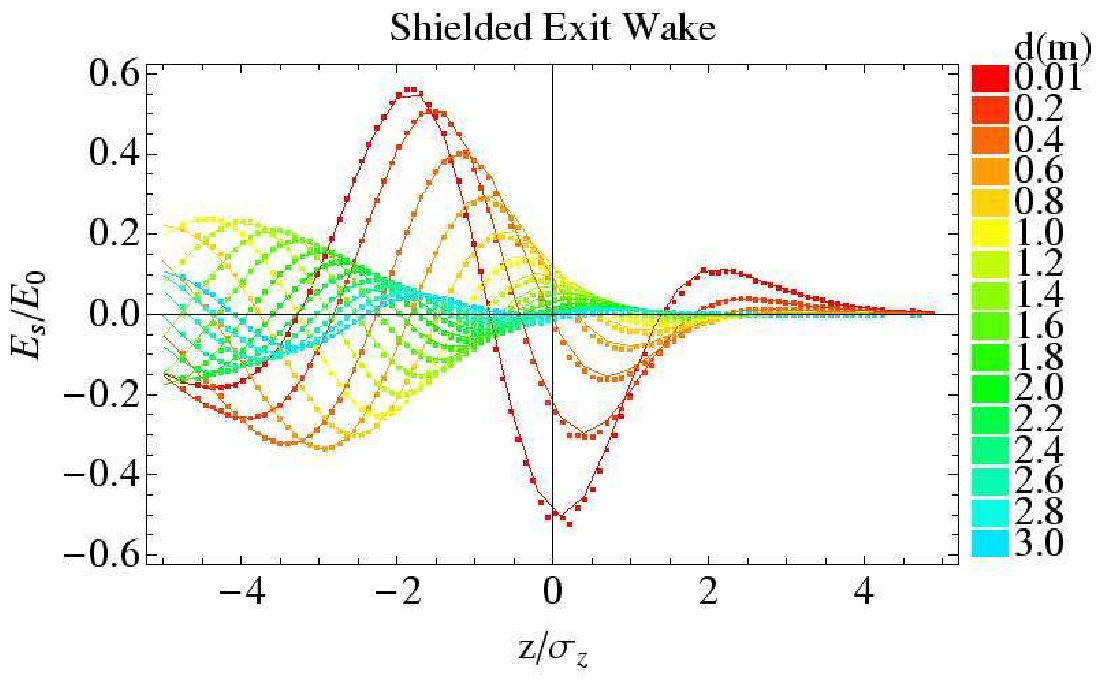}
\caption{Exit wake-field as a function of the length $d$ from the end
of the magnet for free space (top) and with shielding (bottom).  {\tt
Bmad} (dots) shows excellent agreement with CSR-Wake formula
Eq.~(\ref{eq:esexit} (top lines) in the free space case, and with
numerical integration over image bunches using Eq.~(\ref{eq:Eimages})
(bottom lines).  Parameters set A is used.\label{fg:exit}}
\end{figure}

The method in this paper can correctly account for the CSR-Wake in a
drift section following a magnet. Using parameter set A,
Fig.~\ref{fg:exit} shows this wake as a function of the distance $d$
from the end of the magnet in free space and with shielding. The
free space case shows excellent agreement between {\tt Bmad} and the
CSR-Wake formula Eq.~(\ref{eq:esexit}). 

The other codes do not allow simulations with shielding in this
regime. We therefore compare {\tt Bmad} with a numerical solution
using retarded fields. In order to avoid the complication of computing
retarded times and positions of the bunch distribution and its image
charges, we do not use Li\'{e}nard-Wiechert fields, but rather
Jefimenko's equations \cite{jackson99}. In general, the electric field
due to a 1-dimensional charge density $\rho(s,t)$ and current density
$\BfJ(s,t)$ at a position $s$, time $t$, and height $h$ is
\begin{align}
  \BfE (s, t \, ; h ) =  
  \frac{1}{4 \pi \epsilon_0 } &\int ds' \,  
  \Biggl\{ \frac{\BfL}{L^3} \left[\rho(s', t')  \right]  \\
  + &\frac{\BfL}{c \, L^2} \left[
  \frac{\partial \rho(s', t')}{\partial t' } \right] - 
  \frac{1}{c^2 \, L} \left[ 
  \frac{\partial \BfJ (s', t')}{\partial t'}\right] \Biggr\}, 
  \nonumber
\end{align}

where $\BfL$ is the vector from position $s'$ to position $s$ at
height $h$, $L$ is its magnitude, and the brackets $[\, ]$ are
evaluated at the retarded time $t' = t - L / c$. As compared with
integrating over retarded fields at $s'$, using Jefimenko's equations
has the advantage that one never has to solve for the retarded time in
equations of the type $t' = t - |\Bfr-\Bfr(t')|/s$.

The total electric field due to alternating image charges is then

\begin{equation}
\BfE_\textrm{images} (s, t) = 2 \sum_{n=1}^{\infty} (-1)^n \,  \BfE(s,t  \, ; n\, h). 
\label{eq:Eimages}
\end{equation}

Applying this to the geometry of a bend followed by a drift,
Fig.~\ref{fg:exit} (bottom) shows excellent agreement with {\tt Bmad}.

\subsection{Coherent Energy Loss}

\begin{figure}[t!]
\centering
\includegraphics[width=\columnwidth]{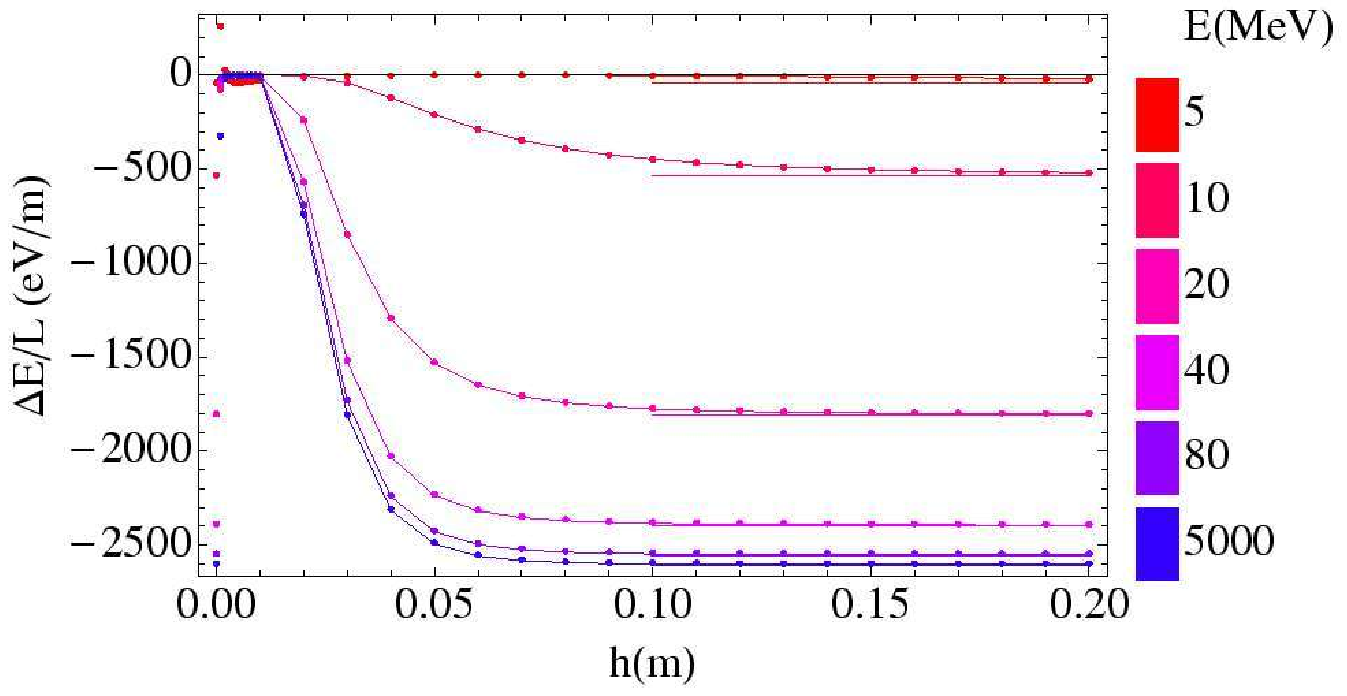}
\includegraphics[width=\columnwidth]{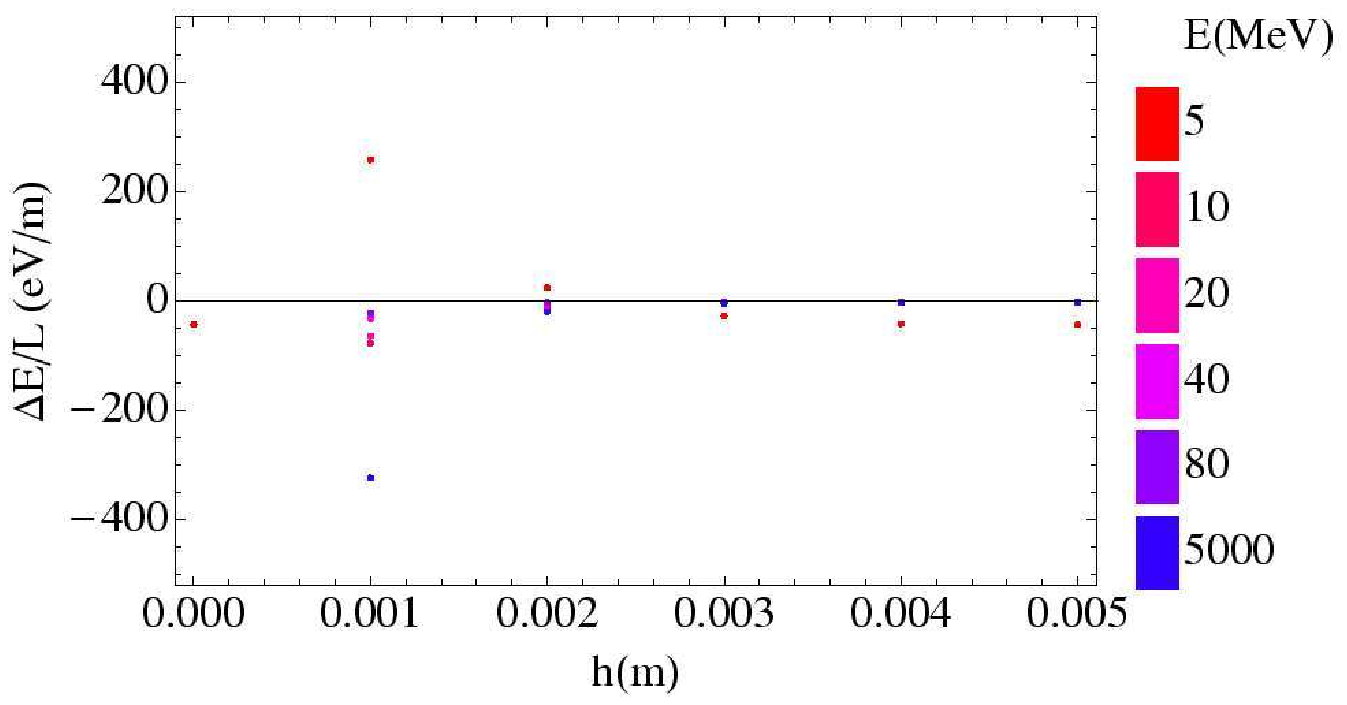}
\caption{Average energy losses versus shielding height for various
energies. The bottom plot is the same as the top highlighting the smaller heights.  {\tt Bmad} (dots) agrees well with numerical integration of
Eq.~(\ref{eq:PNspectrum}) (curved lines) using the shielded power
spectrum of \cite{schwinger45} down to 2mm. Steady state losses
computed using Eq.~\ref{eq:pcoherent} are indicated as horizontal
lines. Parameters set A is used, with the number of image layers $N_i=64$.
\label{fg:e_tot}}
\end{figure}

Our final test of {\tt Bmad} compares the total coherent energy lost
for various particle energies and chamber heights to the integration
of the power spectrum. In general, for $N$ particles traveling on the
same curve at different phases $\phi_n$, the $N$ particle power
spectrum is

\begin{align}
  \frac{d P^{(N)}}{d\omega} &= \left| 
  \sum_{n=1}^{N} e^{i \phi_n} \right|^2 \frac{d P^{(1)}}{d\omega} \\
  &= N\left(1+(N-1)\left| \sum_{m \neq n}e^{i ( \phi_m - \phi_n)} \right|^2 \right) 
  \frac{d P^{(1)}}{d\omega}\nonumber \\
  &\simeq  N  \frac{d P^{(1)}}{d\omega} \label{eq:PNspectrum} \\
  & +N (N-1)\left| \int dz \, \lambda (z) \exp 
  \left(i\, \frac{\omega z}{\beta c}\right) \right|^2 \frac{d P^{(1)}}{d\omega}, 
  \nonumber
\end{align}

where $d P^{(1)}/d \omega$ is the single particle power spectrum, and
$\lambda(z)$ is the longitudinal particle distribution. The first term
in Eq.~(\ref{eq:PNspectrum}) is the incoherent power spectrum, while
the second is the coherent power spectrum. In the presence of
conducting parallel plates, the single particle power spectrum is
given in Eq.~(47) of \cite{schwinger45}. Using this,
Eq.~(\ref{eq:PNspectrum}) can be integrated numerically, and in
Fig.~\ref{fg:e_tot} the resulting coherent part shows excellent
agreement with {\tt Bmad} for energies down to 5MeV and chamber
heights down to 2mm. At smaller heights, the number of image layers
used in the simulation ($N_i = 64$ here) is not sufficient to correctly model the CSR.

In absence of shielding plates, Eq.~(\ref{eq:PNspectrum}) can be
integrated exactly for a Gaussian distribution using the well-known
free space single particle power spectrum.  With a standard deviation
$\sigma_z$, the total power lost by $N$ particles is

\begin{equation}
  P^{(N)} = P^{(1)}N + P^{(1)} \, N(N-1)\, T\left( 
  \frac{3\,  \sigma_z\,  \gamma^3}{2\, R\,  \beta} \right), 
  \label{eq:pcoherent}
\end{equation}

where 
\begin{equation}
  P^{(1)} \equiv \frac{2}{3} r_c m c^3 \frac{\beta^4 \gamma^4}{R^2} 
\end{equation}
 
is the power lost by a single particle, and

\begin{equation}
  T(a) \equiv  \frac{9}{32 \sqrt{\pi} a^3} \exp \left( 
  \frac{1}{8 a^2} \right) K_{5/6}\left(\frac{1}{8 a^2}\right)-\frac{9}{16\,  a^2}.
\end{equation}

This result agrees well with {\tt Bmad} in Fig.~\ref{fg:e_tot}. The
function $T(a)$ can be expanded asymptotically, giving the leading
order coherent energy change in a length $L$ as

\begin{equation}
  \Delta E^{(N)} \simeq -N^2\, r_c m  c^2 
  \frac{\Gamma \left( \frac{5}{6}\right)}{6^{1/3} \sqrt{\pi}}  
  \frac{L}{ \left( R^2\, \sigma_z^4 \right)^{1/3}}.
\end{equation}

This is consistent in the scaling and magnitude of $E_0$ in Eq.~(\ref{eq:e0}). 

\section{Conclusion}

A general implicit formula for the longitudinal kick due to the
coherent synchrotron radiation has been developed for particles on a
common orbit. This formalism will handle any geometry of bends and
drifts. For simulations, this formula is to be preferred over the
explicit ultra--relativistic formula because the implicit formula does
not have a singularity at $z = 0$ and is applicable at lower particle
energies and smaller length scales.

Additionally, a heuristic formula for the longitudinal space charge
kick has been presented which takes into account  transverse
displacements of the kicked particles.

This formalism has been implemented in {\tt Bmad}. We show that the
longitudinal wake field compares well with the A\&Y code of
A\&Y \cite{agoh04} and the CSR-Wake formula of Warnock
\cite{warnock90} for the steady state, with and without CSR shielding
by parallel plates. In the transient case where the A\&Y code
often does not follow the CSR-Wake formula of \cite{derbenev95}
exactly, {\tt Bmad} does agree well with that formula.

\section{Acknowledgment}

The authors would like to acknowledge help from Tsukasa Miyajima and Ivan
V. Bazarov. And we thank Michael Borland for useful discussions. This
work has been supported by NSF cooperative agreement PHY-0202078.

\appendix
\section{Heuristic Shielding Argument}
\label{sc:shielding}

\begin{figure}[ht]
\centering \includegraphics[width=\columnwidth]{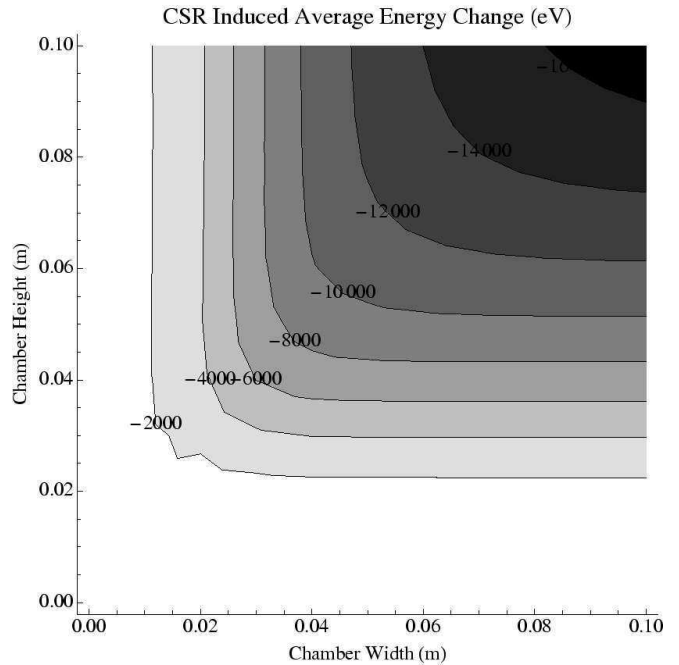}
\caption{Numerical results using the A\&Y code with varying
dimensions of a rectangular chamber.  A 1m, 7.5 degree bend with a 2ps
long bunch of 0.82nC charge was used. Contours represent the energy
change induced due to the CSR wake field in 2000eV increments. The
shielding effect thus not only due to the vertical dimension, but primarily
due to the smaller of the two beam-pipe dimensions.\label{fg:gradient}}
\end{figure}

It initially might seem surprising that vacuum chambers of many
centimeter width can have a shielding effect on synchrotron radiation
with much smaller wavelength. We therefore add a heuristic explanation
here.

Starting with Schwinger \cite{schwinger45}, shielding by a vacuum
chamber has often be considered by studying infinite horizontal,
conducting plates.  But the following heuristic argument indicates why
it is both the vertical and the horizontal boundary that determines
shielding of CSR. Numerically solving Maxwell's equation in the vacuum
chamber supports this argument as shown in Fig.~\ref{fg:gradient}.

A highly relativistic particle emits synchrotron radiation within a
narrow cone. For the radiation's component of wavelength $\lambda$,
the opening angle of this cone is approximately
$\Delta\theta=(\frac{\lambda}{R})^{\frac{1}{3}}$ in the horizontal and
vertical. The opening angle in the vertical determines the vertical
width of the radiation load on the vacuum chamber wall, and the
horizontal divergence can be observed by shining through a pinhole.

The radiation field builds up within a radiation buildup time of
$\Delta t = \frac{R}{c}(\frac{\lambda}{R})^{\frac{1}{3}}$.  During
this time, the radiation fields produced by the electron coherently
add up to form the full radiation power. If the radiation does not
interfere with an obstacle, for example the vacuum pipe, within this
time, the electron looses as much energy as it would without any
vacuum pipe.

The width and hight of the radiation cone that builds up during the
radiation buildup time is therefore given by $w_r\approx h_r\approx
c\Delta t\Delta\theta = R(\frac{\lambda}{R})^{\frac{2}{3}}$. Vacuum
chambers that have smaller dimensions interfere with the radiation
process and shield the part of radiation for which
\begin{equation}
  \lambda \gtrsim \mathrm{Min}
  \left[w\sqrt{\frac{w}{R}} , h\sqrt{\frac{h}{R}} \right]\ .
\end{equation}
Wavelengths are therefore shielded when they are above a length that
is much smaller than the chamber dimensions.

While we have used a very approximate heuristic argument,
Fig.~\ref{fg:gradient} computed by the A\&Y code, indeed shows that
both dimensions can lead to shielding, and that to first approximation
only the smaller of the two dimensions is relevant.


\end{document}